\documentclass[twocolumn]{aa}
\usepackage[utf8]{inputenc}
\usepackage{amsfonts}
\usepackage{amssymb}
\usepackage{amsmath}
\usepackage{natbib}
\usepackage{graphicx}
\usepackage{txfonts}
\usepackage{textcomp}
\usepackage{subfigure}
\usepackage[pagebackref,colorlinks,citecolor=blue,urlcolor=blue,linkcolor=blue]{hyperref}
\usepackage[normalem]{ulem}
\usepackage{cancel}
\usepackage{wasysym}
\usepackage{color}

\graphicspath{{FiguresHP/}}
\newcommand{\degree}{\ensuremath{^\circ}}
\usepackage{siunitx}
\begin{document}

\title{Direct illumination calibration of telescopes at the quantum precision limit}

\author{
E.~Barrelet\inst{1} 
}

\institute{
LPNHE, CNRS-IN2P3 and Universit\'{e}s Paris 6 \& 7, 4 place Jussieu, F-75252 Paris Cedex 05, France
}

\titlerunning{Direct Illumination Calibration of Telescopes at the Quantum Precision Limit}

\offprints{barrelet\@@lpnhe.in2p3.fr}

\date{Received Mont DD, YYYY; accepted Mont DD, YYYY}

\abstract{ The electronic response of a telescope under  direct illumination by a point-like light source is based on photon counting. With the data obtained using the 
SNDICE light source and the Megacam camera on the CFHT telescope, we show that the ultimate precision is only limited by the photon statistical fluctuation, which is below 1 ppm. A key 
feature of the analysis is the incorporation of diffuse light that interfers with specularly reflected light in the transmission model to explain the observed diffraction patterns. 
The effect of diffuse light, usually hidden conveniently in the Strehl ratio for an object at infinity, is characterized with a precision of 10 ppm. In particular, the spatial 
frequency representation provides some strong physical constraints and a practical monitoring of the roughness of various optical surfaces.
}

\maketitle

\section{Introduction}

The calibrated light sources developed for our SNDICE project \citep{ref1} are based on a direct illumination concept (\citealt{ref1'}) using the 
new generation of light emitting diodes (LEDs) to reach a high stability of about 10$^{-4}$. This opens the possibility of measuring the 
sensitivity of good space-based CCD cameras such as those of the Corot and Kepler telescopes with a good precision.
 We have tested the calibration of a telescope and a large-field camera by using the images of the SNDICE light source taken by the CFHT 
telescope equipped with the Megacam camera \citep{ref2}. We placed the limiting precision of the direct illumination calibration of the CFHT 
telescope at the quantum bound ($\approx$10$^{-6}$), which only depends on the potential improvement of the CCD readout electronics described in 
Sect.\ref{sec:4}. Our proof-of-concept goals are distinct from those of the more practical study by \cite{ref3} who used SNDICE to yield an 
improved photometric calibration of the SNLS experiment of about 10$^{-3}$.

A novel feature of the present paper is the comprehensive model of the telescope transmission, including its optical defects. This model, defined 
in Sect.\ref{sec:21}, combines diffuse and specular light in a common photon wave packet (WP) model. It extends the conventional models called here 
"specular models", where the optical surfaces are mathematically defined and the properties of optical media are represented by continuous 
reflection and refraction functions, while the primary light propagation is symbolized by ray optics. Our WP model, which parametrizes the whole 
interference pattern, is validated quantitatively with an exquisite precision.

In Sect.\ref{sec:22}, the spatial frequency spectrum of the interference signal is shown to separate the effect of light propagation in free 
space from that of electronic and optical defects. The former is used as a validation of the model and then taken as a prior. The latter is used 
as an ultra precise control of the optical quality and of the CCD electronic response. Following this spectrum allowed us to monitor during one 
or two hours runs the stability of the interference signal at the quantum precision limit. The only deviation found is due to the microscopic 
motion of the LED source. It is then integrated in the spectral analysis and corrected for. Independently of this analysis of the mirror surface, 
we provide efficient algorithms for detecting, localizing, and parametrizing the defects of the Megacam camera in Sect.\ref{sec:25}. This is a 
first step, since there are about 10$^{5}$ such defects to monitor individually during the life of the camera. Their effect on 
astronomical images is obviously diverse and cannot be represented by a simple pixel-to-pixel correction.

The last part of our analysis describes the successive steps of a complete telescope photometry based purely on photon counting.  This analysis 
implicitly uses the spectral properties of the interference pattern found in Sect.\ref{sec:22} and the mitigation of the electronic problems 
found in Sect.\ref{sec:4}. First Sect.\ref{sec:51} defines the four operators (pixel combinations) that permit a clean photon counting 
analysis and introduces their pure Gaussian properties that allow precisely applying the law of large numbers up to the 10$^{13}$ photons 
contained in a Megacam image. Second, Sect.\ref{sec:52} and Sect.\ref{sec:54} establish the second-order corrections to the pure Gaussian model 
needed for multinomial statistics and for LED motion checks, respectively. Last, Sect.\ref{sec:55} and Sect.\ref{sec:56} apply these methods 
to flux and noise estimation respectively. Combining these two methods, we show that the fluctuation of pixel counts is rigorously proportional to the 
square root of the flux.
   
More generally, these methods offer a perspective to replace the paradigms of the classical optics and the photometric standards by paradigms relying 
on fundamental physics. The technical breakthrough behind these progresses, beyond the new optical sources and the detectors already mentioned, is 
clearly the data processing power which is essential for the analyses presented in this paper.

\section{Using coherent light for calibration}
\label{sec:2}

\begin{figure}
\centering
\includegraphics[width=1.0\linewidth]{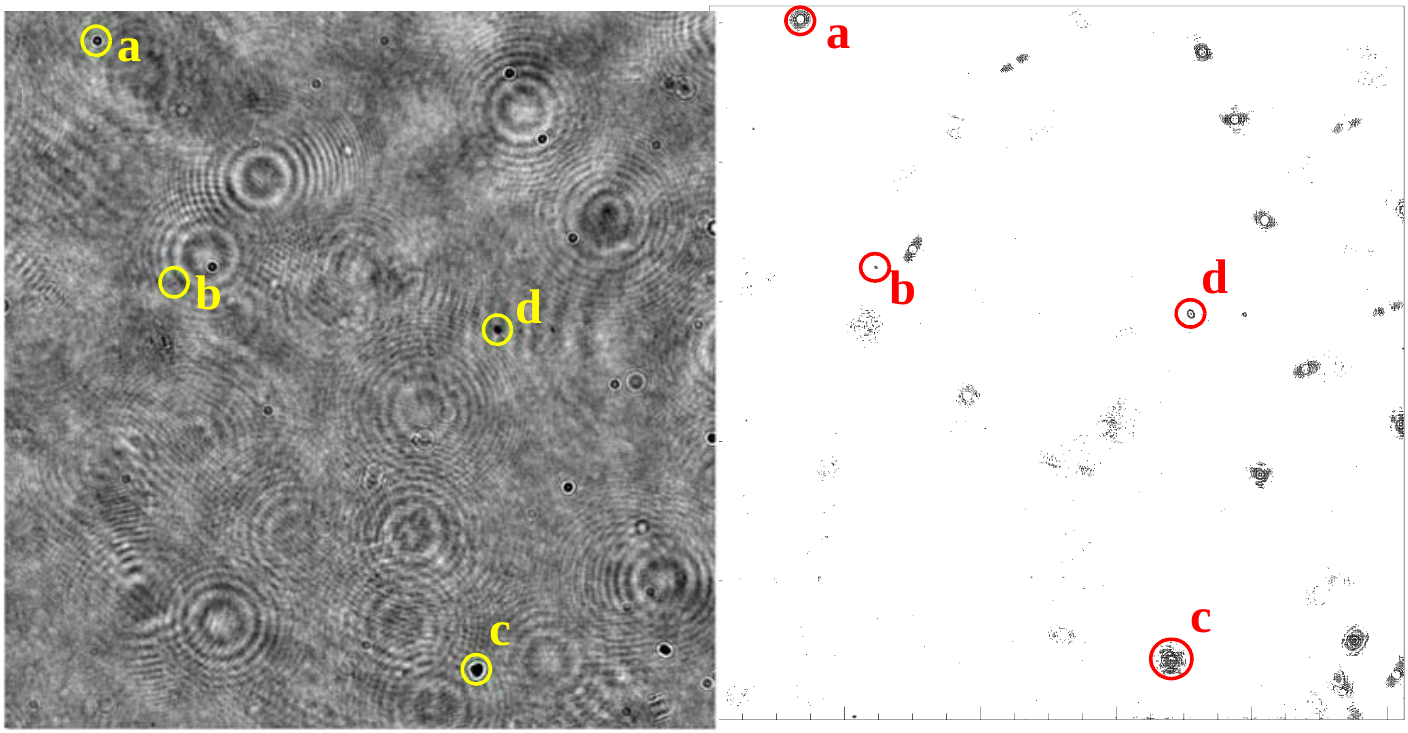}
\caption{{\bf a) (left)} Wave packet signal (WP) measured in a 1024$\times$1024 pixel$^2$ area. The gray scale covers <WP> $\pm$1.5$\sigma$.
{\bf b) (right)} The 1\% of pixels in this area with a sharp WP gradient due to defects of the camera optics 
($\|\protect\overrightarrow{\vec{\nabla (WP)}}\| \geq 16 \sigma$). Four camera defects a,b,c,d of different types are circled in the two figures for 
discussion in Sect.\ref{sec:25}.
 \label{fig:fig2}}
\end{figure}

\begin{figure} \centering \includegraphics[width=1.0\linewidth]{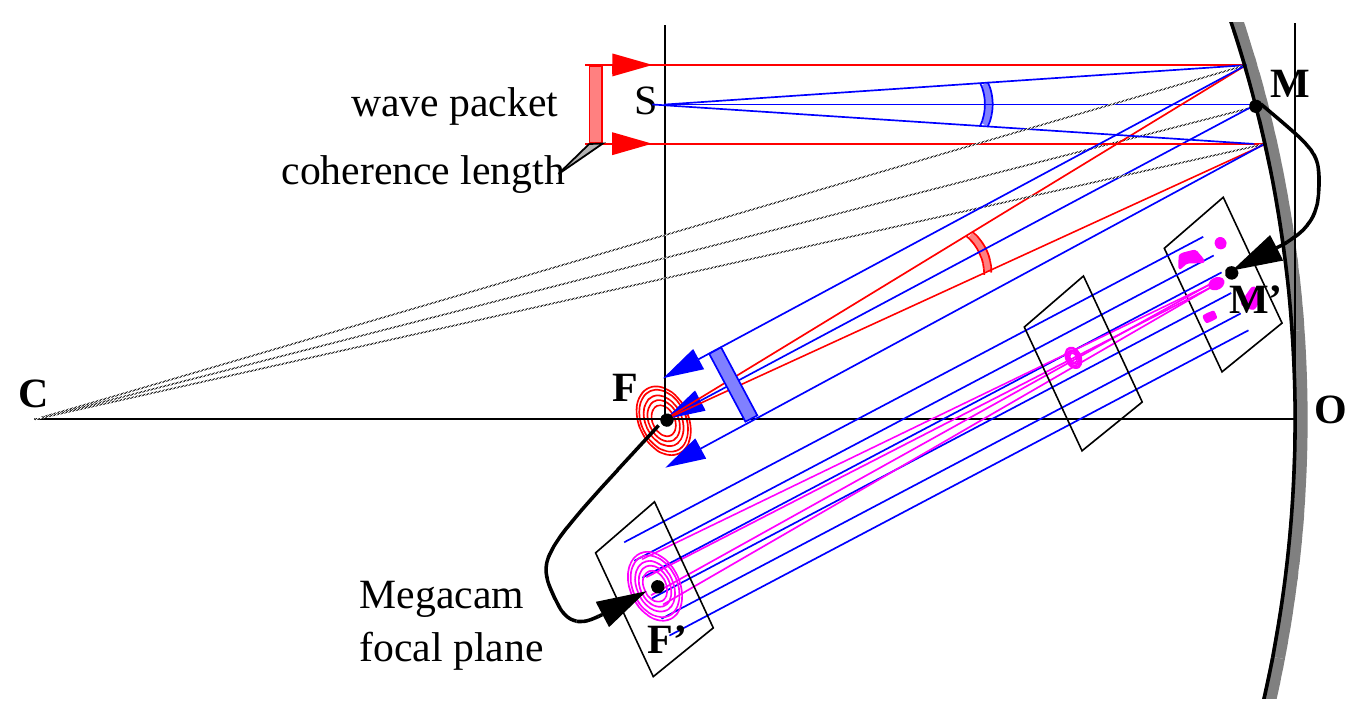} \caption{Ray optics and wave packet: in red we plot a star 
source at infinity imaged at focus F and its diffuse reflection falling in the Strehl ratio area around F; in blue we show a LED 
source S at focal distance and diffuse reflection interfering with specular reflection (reflected beam MF displaced for clarity into 
M'F').
 \label{fig:fig1}}
\end{figure}

Measuring the overall response of a telescope by placing a point illumination source (i.e. a partially coherent source) at the focal distance is 
attractive because it is expected to yield smooth images, and each pixel of the camera would define a single light ray. Previous attempts 
\footnote{cf. Stubbs, C. et al. (unpublished)} have met the obstacle also seen by SNDICE (Fig. \ref{fig:fig2}.a), which is a plethora of diffraction 
patterns that is due to the imperfections in the mirror surface. We can consider these artifacts as a nuisance caused by the partial coherence, 
but they alter an image exactly as they would for a target object at infinity, as suggested by Fig. \ref{fig:fig1}, based on the classical 
Fraunhofer diffraction theory (\citealt{ref0}, chapter VIII, fig 8.6). Therefore they need to be taken into account by astronomical 
calibration. The first goal of this section is to demonstrate the exact correspondance of the light diffracted by the same area of the mirror, 
either from a point source at infinity or at a focal distance. Diffracted light is expressed by the Fresnel diffraction integral as a convolution 
product of an aperture function representing the defects of the mirror and the impulse response of the free space propagation from the mirror to 
the focal plane. This property is used in Sect.\ref{sec:22} to extract by Fourier analysis pure diffracted light from non-diffracted light. In 
Sect.\ref{sec:23} we measure the effect of the translation of the point source on the diffracted light. By joining these two developments, we can 
compare extended source and point source images. We can also measure the stability of the SNDICE source with high precision.

\subsection{SNDICE geometrical setup and the generation of a CCD image }
\label{sec:21}

In our setup drawn in Fig. \ref{fig:fig1}, the SNDICE LED source in S is a 0.1 mm$^{2}$ chip situated at a focal distance f=13.5 m (19 
m in reality) from the mirror. The SNDICE axis SM is aligned with the telescope axis FO. It pierces the mirror in M within its 1.8 m 
radius. The SNDICE beam one degree aperture just covers focal plane (no stray light). The Megacam camera is centered on the focus F of 
the parabolic mirror. A pixel covers a 13.5$\times$13.5 $\mu m^{2}$ area, that is, a 1.0 $\mu rad^2$ solid angle. The center of 
curvature C of the mirror is used for geometrical ray tracing. The optics is completed by an image corrector, made of four lenses and 
one out of five filters. Filters are not used in this study. The optical band-pass provided by a LED spectrum is $\Delta \lambda 
/\lambda\approx$5\%, a third of the band-pass of a typical astronomical filter.

The base concept is to consider the photon transmission through the telescope as a quantum process as well as the photon emission 
(LED) and the photon absorption (CCD). Telescope calibration establishes the balance sheet between a photon-counting calibrated light 
source and a photon-counting light detector. The LED emits a thin spherical wave packet (WP). Planar after reflection on the mirror 
around M, the wave packet collapses when absorbed by the focal plane in F. The WP probability of counting a photon in a given CCD pixel 
${\{v\}}$ is  \begin{equation} P b \{ \gamma \in S \rightarrow e \in v \} =|\psi|^2 \ast \delta_{\{ v \}} = A_{\{ v \}}
 \label{eq:1}
\end{equation}
where |$\psi$(x,y)|$^2$ is the modulus square of the wave function amplitude and $\delta$(x,y)$_{\{v\}}$ the electron collection efficiency.
The total number of photons falling on the channel $k$ is
\begin{equation}
\underbrace{N^{adu}_{i,j,k,I}}_{N\{v\}} = \underbrace{\phi_I \times \Delta t \times L(T)}_{\Phi_I} 
\times \underbrace{a_{i,j,k,I} \times b_k}_{A\{v\}} 
\times \underbrace{\epsilon_ {k'} \times g_{k,I}}_{g\{v\}}
+ \underbrace{P_ {i,j,k,I}}_{P\{v\}} \quad\quad . 
\label{eq:2}
\end{equation}

Assuming an uniform photon emission rate $\phi_I$, the expected photon total count $\Phi_I$ during the exposure $I$ is proportional to 
the exposure duration $\Delta t_I$, with a temperature correction $L(T_I$). The photoelectron count in one pixel follows a Poisson law 
with an expected value $\Phi_I \times$A$_{\{v\}}$. The counting rates in all individual pixels or any subset inside the complete 
3.4$\times$10$^{8}$ pixel set $\{v\}$ follows a multinomial law.

The last part of the Eq. (\ref{eq:2}) represents the digitization of the pixel counts, represented ideally by two constants: a gain 
factor g$_{\{v\}}$ that transforms a number of photons into a number of ADUs (analog to digital unit) and the pedestal P$_{\{v\}}$. 
These electronic constants are two Gaussian variables whose fluctuations have to be added to the multinomial fluctuations of the 
photon counts to constitute the global statistical factor studied in Sect.\ref{sec:56}. Both electronic constants are studied 
specially in the electronic section Sect.\ref{sec:4}, but we state here that they have defects that introduce strong 
variations (1.5$\times10^{-3}$ RMS) from one image $I$ to the next, depending on the electronic 
channel $k$. For this reason, these constants are indexed with $I$ and $k$ instead of being considered as long-term constants. We take 
into account that the gain fluctuation equally affects all the pixels read by one channel $k$ and that the quantum 
efficiency $\epsilon_k$ is the same for the two channels inside one CCD. For this purpose, Eq. (\ref{eq:2}) introduces the fraction 
$b_k$ of the total number of photons falling on the channel $k$ and the respective quantum efficiency $\epsilon_k'$. The image matrix 
$a_{i,j,k,I}$ depends on $I$ because the interference pattern slides (LED jitter), which we extensively study in 
Sect.\ref{sec:54} and which is normalized by $\sum\limits_{i,j} a_{i,j,k,I} = 1$.

\subsection{Wave packet signal and its Fresnel spectrum }
\label{sec:22}

\begin{figure} \centering \includegraphics[width=1.0\linewidth]{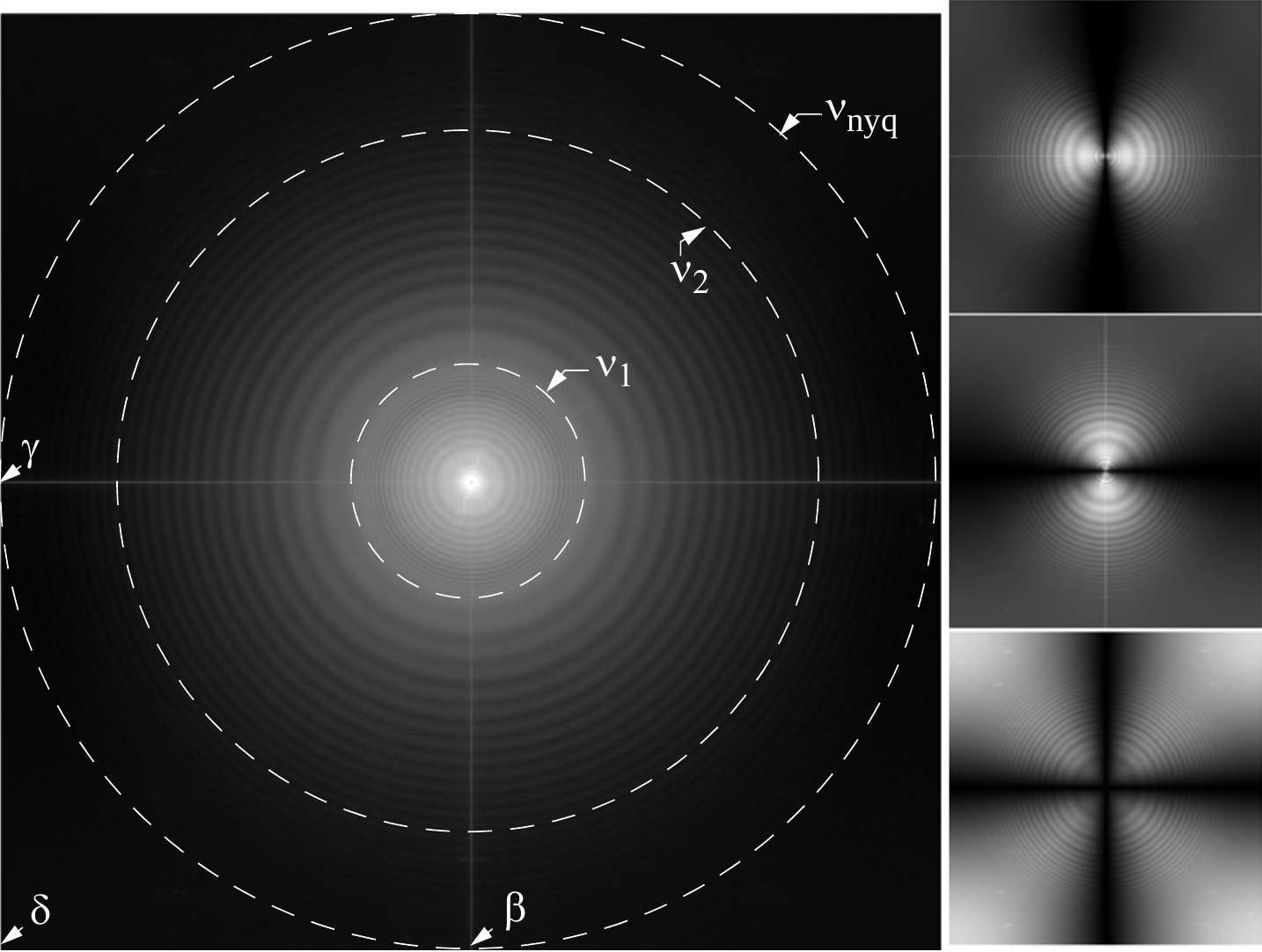} 
\caption{Effect of the Fourier transform on a 1024x1024 vignette of the Sndice image. {\bf left:} applied to the original field; {\bf 
right:} applied to the 3 PDE transformed fields -$\nabla_x$, $\nabla_y$, $\Delta^2/\Delta x\Delta y$-. The 288 vignettes covering the 
focal plane are quadratically averaged. The frames are set by |$\nu_x$| and |$\nu_y$|=$\nu_{nyq}$. The horizontal and vertical lines 
passing through the center are due to electronic noise. Dashed circles mark radial cuts $\nu_1$ and $\nu_2$ in Fig. \ref{fig:fig4_a}. The 
three points $\beta$, $\gamma$ , and $\delta$ mark the three real FFT components averaging same name filters ($\alpha$ is at the center).
 \label{fig:fig3}}
\end{figure}

Within the quantum mechanical framework, each individual photon wave function carries the complete interference pattern, and the image 
builds up by independently piling up a large number of photoelectrons (10$^{12}$/s) in all pixels. Accordingly, the optical modulation 
of an image is perfectly represented in Eq. (\ref{eq:1})\&(\ref{eq:2}) by a probability density, constant at a 10$^{-6}$ precision 
level for hours. Before proving it in Sect.\ref{sec:56}, we show here that the wave packet signal conforms to the laws of optics. Our 
LED light propagates in free space excepted for the reflection on the mirror surface, which can be represented by a Fresnel integral. 
(We neglect the diffraction on the optical surfaces of the image corrector optics, which is treated separately in Sect.\ref{sec:25}). 
We subdivided the mirror into sections covered by some 1024$\times$1024 pixel sub-matrices. The WP signal in each section (e.g., in Fig. 
\ref{fig:fig2}-a) was then Fourier transformed. The Fresnel integral being a convolution of the Fresnel free space propagation function 
and a mirror defect distribution, its transform is the product of two terms: the well-known Fresnel diffraction figure, and the 
transform of the mirror defect distribution. We call the distribution of the modulus the Fresnel spectrum. The quadratic average of 
all spectra is shown in Fig. \ref{fig:fig3}(left).

The spectrum is contained in a square defined by spatial frequencies |$\nu_x$| and |$\nu_y$|$\leq \nu_{nyq}$. The Nyquist frequency 
$\nu_{nyq}$ is 37 mm$^{-1}$, that is, the inverse of twice the pixel width 1/(2$\times$ \SI{13.5}{\micro\metre}). The spectrum, being 
the digital Fourier transform (DFT) of a real function, is centrally symmetric. The rotational invariance around the center ($\nu_x$=$\nu_y$=0) is predicted by 
Fresnel symmetry.

We explain the two lines on the $x$ and $y$ axes crossing at the center by residual electronic problems\footnote{after a large 
reduction by mitigation of main electronics problems} such as pixel-to-pixel (e.g., dead column) or line-to-line (e.g., microphonic 
noise), respectively. The bright spot at the center is due to specular reflection, which is in this way separated from diffraction.

\begin{figure}
\centering
\includegraphics[width=0.8\linewidth]{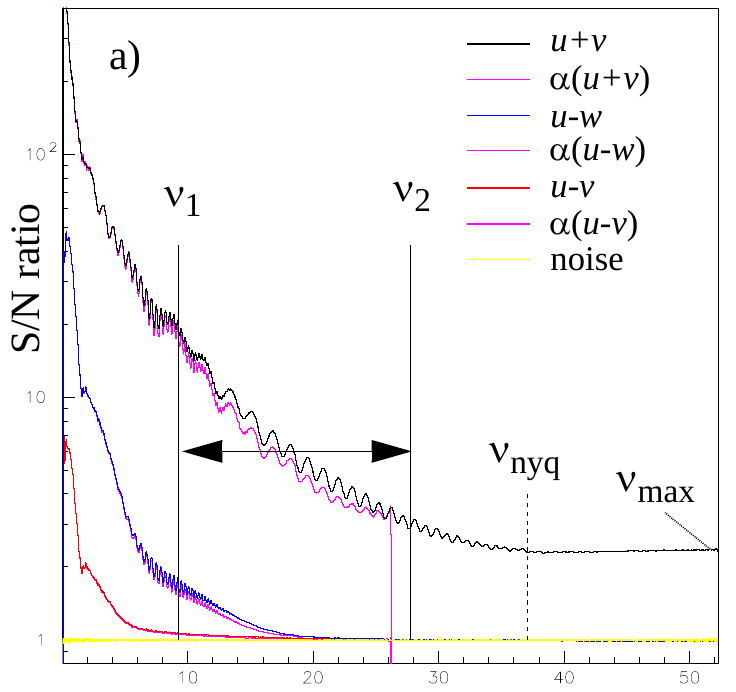}
\caption{Radial spectra of the sum and the difference of two images (image $v$ is close to $u$; $w$ distant). All are normalized to 
the 0.7\% rms photon noise. The [$\nu_1, \nu_2$] cut yields the angular plots in Fig. \ref{fig:fig5}. The Nyquist frequency is $\nu_{nyq}$.
 \label{fig:fig4_a}}
\end{figure}

\begin{figure}
\centering
\includegraphics[width=1.0\linewidth]{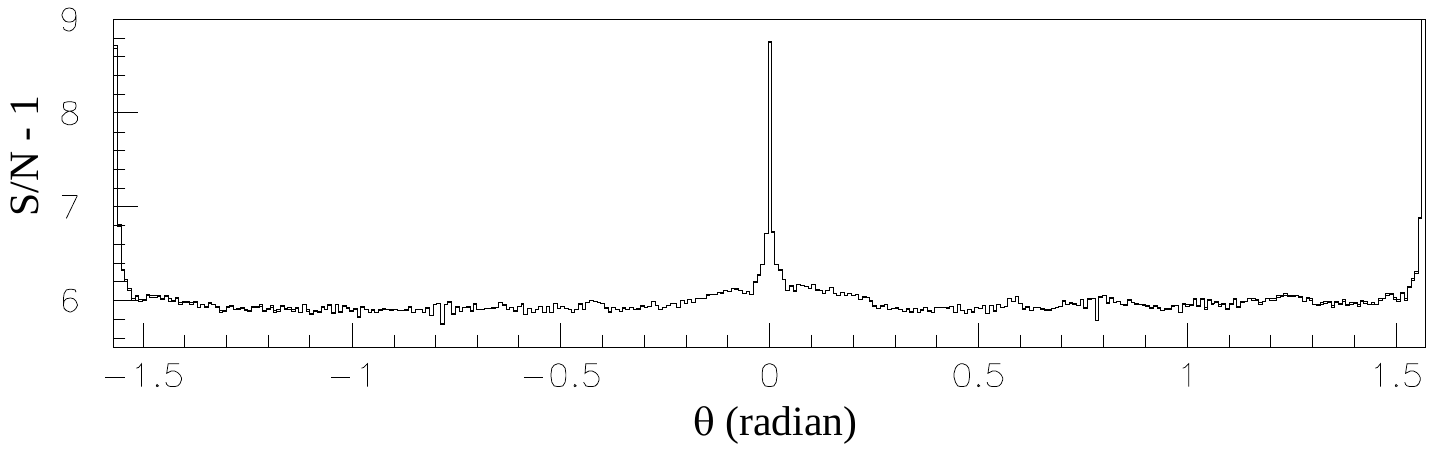}
\caption{Spectral angular distribution of $u$+$v$, normalized by the photon noise and then by subtracting 1. The angular average 
S/N$\approx$6 equals the radial average inside the [$\nu_1$, $\nu_2$] cut. The electronic noise peaks at $\theta$=0 and $\theta$=$\pm \pi/2$.
 \label{fig:fig5}}
\end{figure}

The rotational invariance of the DFT field reduces the amount of empirical data representing the surface state of the mirror by a huge 
factor. Instead of a two-dimensional spatial frequency plot such as Fig. \ref{fig:fig3}, we can make two one-dimensional spectral 
curves: one radial frequency distribution (Fig. \ref{fig:fig4_a}), and one angular distribution (Fig. \ref{fig:fig5}).

Each sample of the field in the spatial frequency plane $\{\nu_x,\nu_y\}$ is taken as a sample at the radial frequency $\nu_{\rho}$ 
and the angle $\theta$~:
\begin{equation}
\nu_{\rho} = (\nu^2_x + \nu^2_y)^{1/2}  \qquad \theta = atan(\nu_y / \nu_x)  \quad\quad .
 \label{eq:6}
\end{equation}

The samples can either be integrated in angular and radial bins, or averaged\footnote{take a white-noise CCD image (photon flat 
field): its |DFT| field is flat. The integrated radial spectrum, proportional to the surface in a given ring, rises linearly with 
radius and then decreases when the rings are no longer contained in the square. The averaged radial and angular spectra are flat, as 
seen in Fig. \ref{fig:fig4_a} and Fig. \ref{fig:fig5}.}. We remark that the radial sampling along a diagonal is defined by a spacing 
divided by $\sqrt{2}$ and that the square pixel sampling filter projected on a diagonal is a triangle with a 13.5$\times \sqrt{2}$ 
micron base. The highest radial sampling frequency $\nu_{max}$ is $\sqrt{2} \nu_{nyq}$ ($\nu_{max}$=52.4mm$^{-1}$). The radial 
spectrum corresponding to the field in Fig. \ref{fig:fig3} is found in Fig. \ref{fig:fig4_a}. Two images are added to this spectrum. 
This allows comparing in the same figure the radial spectrum of the sum of two images (in black) with their difference (blue and red). 
For a sum there is no effect depending on the choice of the images. In contrast, for a difference there is an effect that is related 
to the vicinity of the images in time. There is a greater difference between the images $u$ and $w$ taken after waiting for one hour 
(blue) than between u and v taken within a one-minute delay (red). This effect is explained in Sect.\ref{sec:54} by a progressive 
drift of the LED position with respect to the optical axis of the telescope.

The angular distribution of the sum spectrum, seen in Fig. \ref{fig:fig5}, is computed within the ring $\nu_1<\nu_{\rho}<\nu_2$ . 
The two-dimensional Fourier transform of a real function being centrally symmetric, we need to plot only one half of the unit circle 
$(-\pi/2<\theta <\pi/2)$. As predicted by our model, the distribution is flat, except for the electronic noise, which yields 
accumulations and peaks at $\theta$=0 and $\theta \pm \pi/2$ (where line or column frequencies are null).
 
\begin{figure}
\centering
\includegraphics[width=1.0\linewidth]{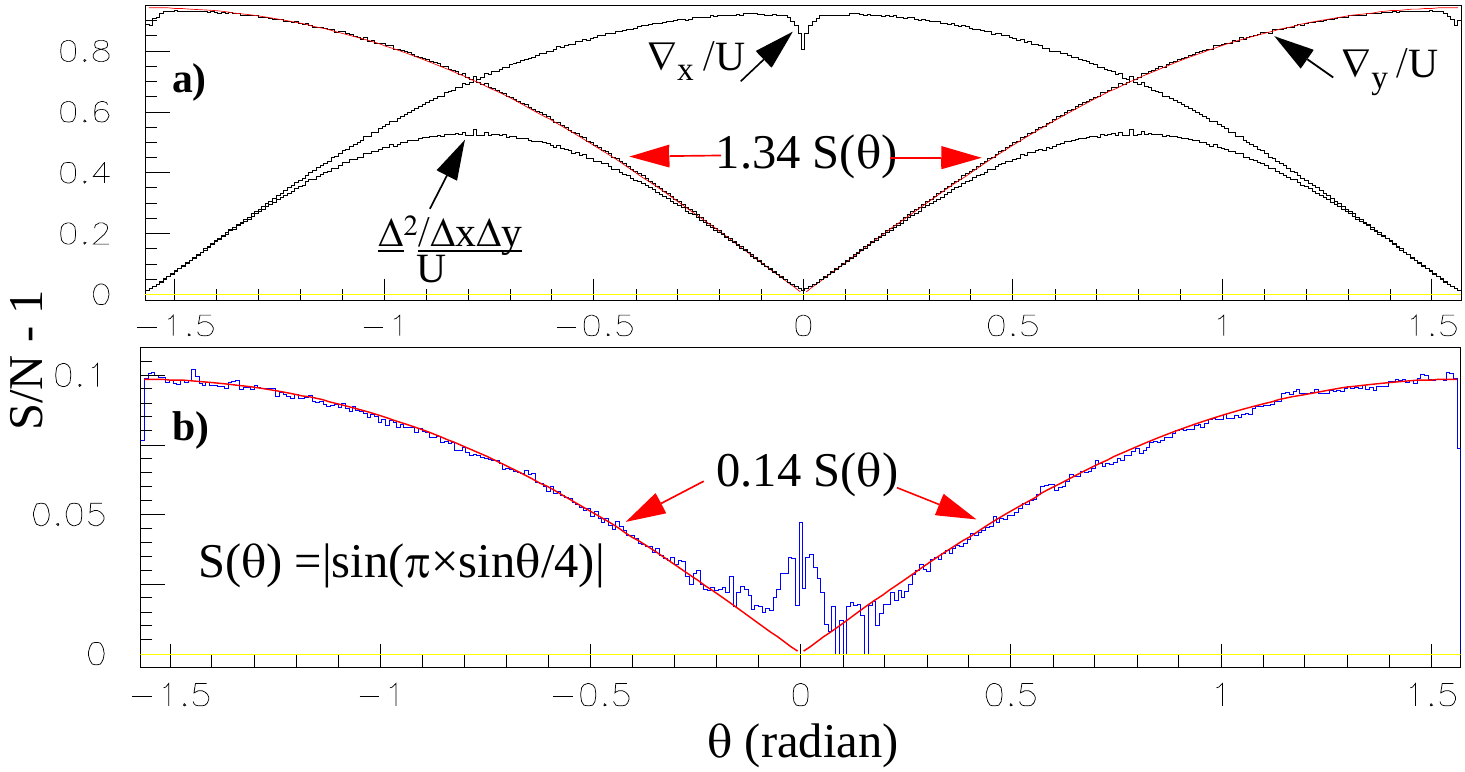}
\caption{ {\bf a)} Spectral angular distribution of $u$+$w$ after applying PDE operators, divided by the unity distribution U of Fig. 
\ref{fig:fig5}. {\bf b)} Spectral angular distribution of $u$-$w$, divided by U (inside the radial cut [$\nu_1$, $\nu_2$]). The 
uncorrelated random spectra, null after subtraction of unity, are plotted in yellow. Fitted curves are shown in red.
 \label{fig:fig6}}
\end{figure}

A Fresnel spectrum is a stable and reproductible characteristic of the status of a section of mirror (2 cm$^2$ (one CCD vignette), 0.1 m$^2$ 
(one SNDICE image), or 8 m$^2$ (the whole mirror)). A complete mirror scan lasts about two hours. Systematic studies such as aging or color 
dependence have not been made so far.

\subsection{Effect of a transverse LED motion on the Fresnel spectrum}
\label{sec:23}

Figure \ref{fig:fig1} shows that moving the source S moves its projection M, the center of the illuminated section of the mirror, but 
keeps it projected on the center of the focal plane. We call $T_x$ and $T_y$ the two translation operators representing a shift of 
the mirror points reflected on a given pixel by one pixel leftward or upward. Partial derivatives of the WP signal are approximated by 
the finite differences of the translation operators $T_x$ and $T_y$ and the identity operator U,

\begin{equation}
 \nabla_x =T_x -U ;\, \nabla_y=T_y -U ;\,
 \Delta^2/\Delta x \Delta y =(T_x -U)(T_y -U) \quad .
 \label{eq:3} 
\end{equation} 

When we apply the these operators, commonly named gradient and hessian partial derivative equation (PDE) filters, to a CCD image, we obtain
three rather uniform new images.  The Fresnel spectra of these images are seen on the right of the main spectrum in Fig. 
\ref{fig:fig3}. Simple mathematics predict the shape of these images. For instance, the DFT of the translation $T_y$ yields the product 
of the complex DFT by a phase shift factor exp(i$\pi \nu_{y} / \nu_{nyq}$). Subtracting unity and taking the modulus gives the 
observed result: the two-dimensional spectrum of $\nabla_y$ is the whole spectrum multiplied by a sin($\pi \nu_{y} / 2\nu_{nyq}$) factor. The 
$\nabla_x$ formula is obtained by exchanging x and y and $\Delta^2/\Delta x \Delta y$ by multiplying the two angular factors.

The angular spectra resulting from the application of PDE filters to the sum of the images are found in Fig. \ref{fig:fig6}.a). They are 
explained by the factor introduced in DFT by differentiation. For $\nabla_y$, the factor is |$sin(\pi \nu_y /2\nu_{nyq})$|. Inside the 
ring $\nu_1<\nu_{\rho}<\nu_2$ the average value of $\nu_{\rho}$ is $\nu_{nyq}$/2 and $\nu_y = \nu_{\rho}\times$ sin $\theta$, therefore 
the angular factor is:

S($\theta$)=sin($\pi$sin$\theta$/4)

The $\nabla_y$ spectrum in Fig. \ref{fig:fig6}.a) and the spectrum of the image $u$-$w$ in Fig. \ref{fig:fig6}.b) are both proportional to 
S($\theta$). A fit yields the respective factors 1.34 and 0.14. For the other pair of images $u$ and $v$ taken at one-minute 
intervals, the angular spectrum is almost null.
 This proves that a LED drift in the y direction is the cause of the small difference between exposures $u$ and $w$. When we apply a 
proportional rule of thumb, the u-w shift distance is a tenth of that of a one CCD line shift computed in $\nabla_y$ (13.7$\mu$). Hence 
we estimate a 1.4$\mu$ LED shift in one hour!

\subsection{Orthogonal basis of differential operators: $\alpha, \beta, \gamma$, and $\delta$}
\label{sec:24}

Similarly we introduce four orthogonal operators $\alpha, \beta, \gamma,$ and $\delta$ which have a crucial role in our image analysis method:

\begin{equation}
\begin{split}
\alpha =(T_x +U)(T_y +U) \quad;\qquad \beta =(T_x -U)(T_y +U) \quad; \\
 \gamma =(T_x +U)(T_y -U) \quad;\qquad \delta =(T_x -U)(T_y -U) \quad\quad .
\end{split}
 \label{eq:4}
\end{equation}

We modified these operators to project the original CCD images $\{a_{i,j}\}$ into lower resolution images (scale 1/2$\times$1/2) by 
restricting indices to even values. More explicitly, we developed Eq. (\ref{eq:4}) using the pixels $a_{i,j}$ defined in Eq. (\ref{eq:2}):

\begin{equation}
\begin{bmatrix}
\alpha_{m,\ n} \\
\beta_{m,\ n}  \\
\gamma_{m,\ n}  \\
\delta_{m,\ n}  \\
\end{bmatrix}
= \frac{1}{2} 
\begin{bmatrix}
1  &   1 & 1  & 1  \\
1  &   1 & -1 & -1 \\
1  &  -1 & 1  & -1 \\
1  &  -1 & -1  & 1  \\
\end{bmatrix}
\begin{bmatrix}
a_{2m+1,\ 2n+1} \\
a_{2m+1,\ 2n}  \\
a_{2m,\ 2n+1} \\
a_{2m,\ 2n} \\
\end{bmatrix} \quad\quad .
 \label{eq:5}
\end{equation} 

The operator $\alpha =\{ \alpha_{m,\ n}\}$ sums pixels with adjacent even and odd indices. Its spectrum in Fig. \ref{fig:fig4_a} follows 
the original spectrum ($U$), but stops at half the frequency range.

$\beta$ and $\gamma$ are similar to the gradient $\overrightarrow{\nabla}$= $\{\nabla_x, \nabla_y\}$, and $\delta$ to the Hessian 
$\Delta^2/\Delta x \Delta y$. Taking the four $\alpha, \beta, \gamma,$ and $\delta$ images together, we have an efficient lossless 
encoding of the original image, represented by the orthogonal matrix of Eq.  (\ref{eq:5}).

\begin{figure}
\centering\includegraphics[width=0.8\linewidth]{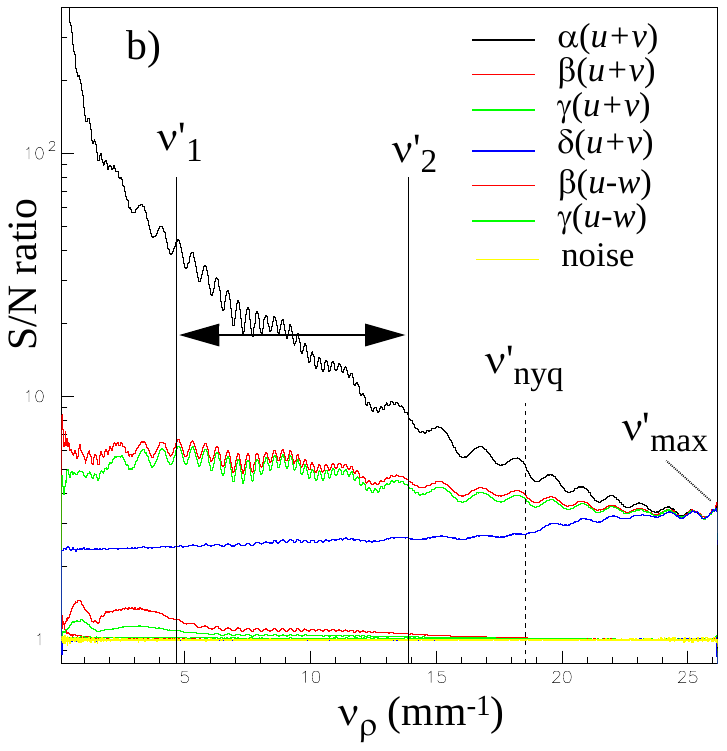}
\caption{Radial spectra of the sum and difference of two images after the application of four filters ($\alpha, \beta, \gamma, \delta$). 
 \label{fig:fig4_b}}
\end{figure}

In Fig. \ref{fig:fig4_b} we show the radial spectra of $\alpha, \beta, \gamma$, and $\delta$ for image sums and differences. The spectra 
of the image differences are almost drowned in the noise, except for those of $\beta$ and $\gamma$ for distant images $u$ and $w$. The 
radial spectra of the $\beta, \gamma$, and $\delta$ operators are cut severely at low frequencies, but higher frequencies are unchanged. 
The four radial spectra converge at $\nu_{max}'=\nu_{max}/2$.

\begin{figure}
\centering
\includegraphics[width=0.8\linewidth]{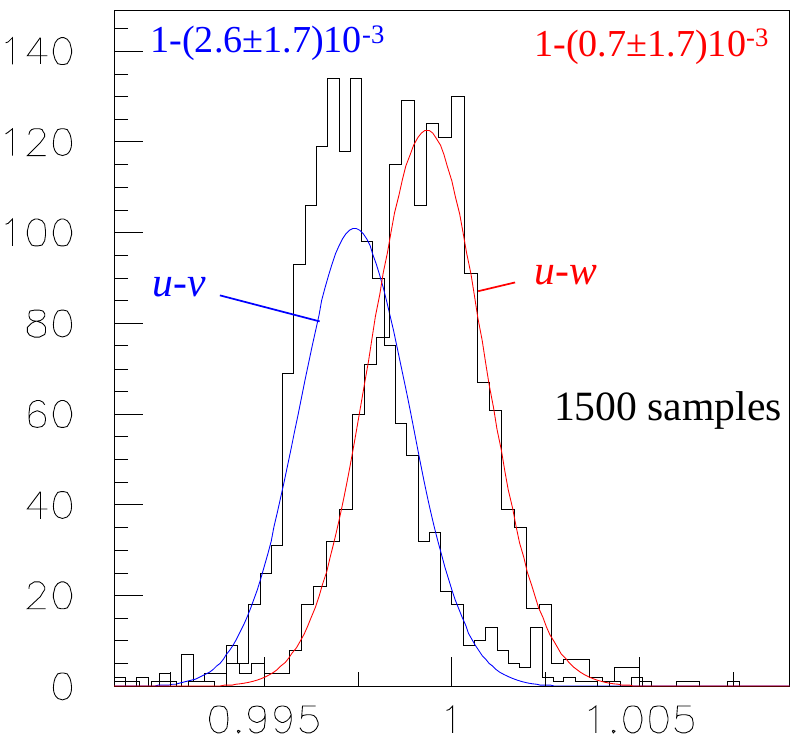}
\caption{Residuals of $\delta$($u$-$v$)/noise and $\delta$($u$-$w$)/noise linear fits of radial frequency spectra (noise=98 adu). 
The precision for each sample is $\approx$10$^{-5}$, and for the whole image the average is $\approx$0.3x10$^{-6}$.
\label{fig:fig7}}
\end{figure}

The uncorrelated photon noise spectrum was obtained by simulation, using a Gaussian variable generator\footnote{adjusted in the residual 
plot in Fig. \ref{fig:fig7}} with an rms equal to 98 adu for each pixel, which corresponds to about 4$\times$10$^4$ photon/pixel. It was 
processed in the same way as for a real image. The resulting radial and angular spectra are flat for the $\alpha, \beta, \gamma$, and 
$\delta$ operators but not for the PDEs. We tested with the highest precision defined by the photon statistics of whole images the 
hypothesis that the $\delta$ operator applied to a difference of images yields a pure photon noise spectrum. For this we fit a flat 
$\delta$ radial spectrum on the $u$-$v$ and $u$-$w$ images. The histograms of the residuals are shown in Fig. \ref{fig:fig7}. The 
reference level of 1 corresponds to the approximate level of the noise (98 adu). The dispersion of the radial samples is 0.17 adu (rms) 
on a mean signal of 17000 adu, that is, $\approx$10$^{-5}$. The number of photons contributing to one sample is the whole content of two 
images: 56$\times$10$^{12}$ divided by 4 (the number of estimators) and then by 1500 (the number of samples), that is, $\approx$10$^{10}$. 
This verifies that the dispersion of the radial samples (10$^{-5}$) is consistent with photon statistical error (10$^{10}$)$^{-1/2}$.

 \subsection{Detection of camera defects}
\label{sec:25}

The photon propagation from the camera lens to the focal plane is shorter than from the mirror by a factor larger than 15. Therefore its 
Fresnel spectrum is much sharper. We developed a test on this premise (\citealt{ref4}), using the gradient vector length:

\begin{figure}
\centering
\includegraphics[width=1.0\linewidth]{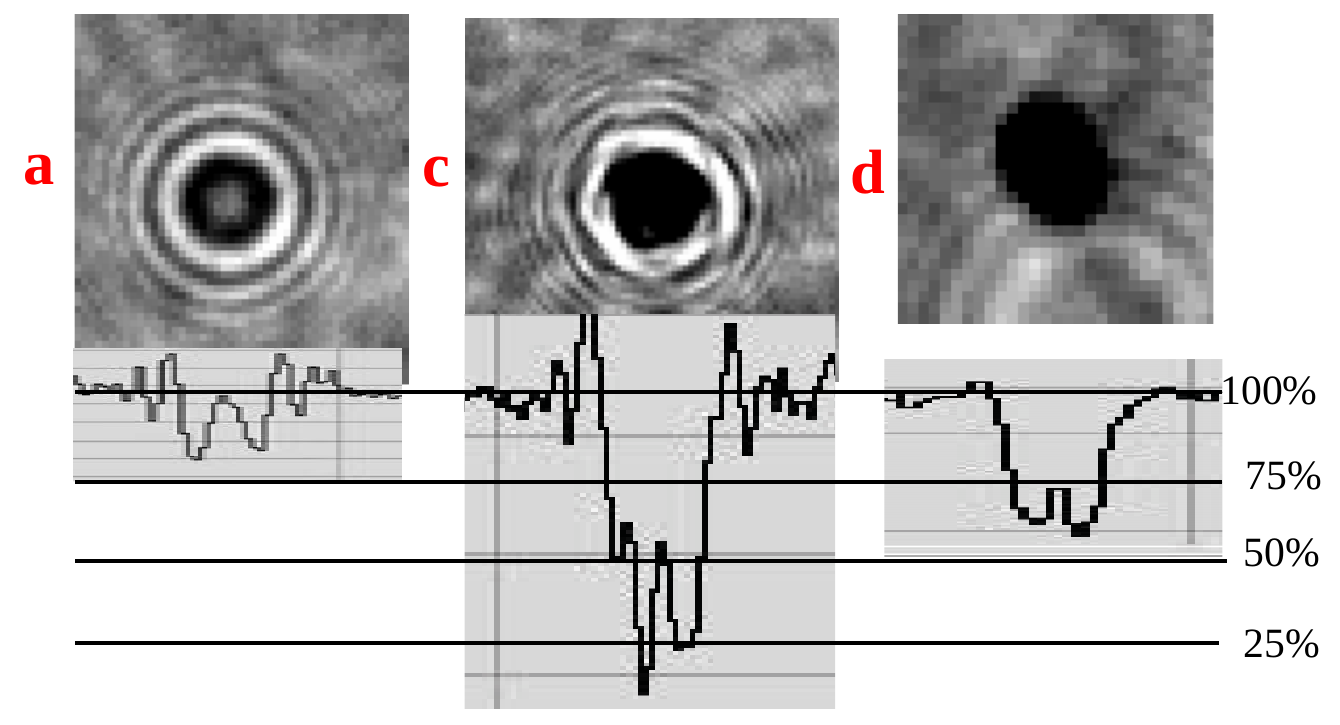}
\caption{Camera defects a, c, and d circled in Fig. \ref{fig:fig2}. {\bf Top:} the flux map around the defects.
{\bf Bottom:} the profile of the adc content drawn along a line passing through the center of the defects.
 \label{fig:fig8}}
\end{figure}

\begin{itemize}
\item[$\bullet$] 1\% of the pixels are tagged by the ||$\overrightarrow{\nabla}$||$\geq$16$\sigma$ cut (see Fig.\ref{fig:fig2}b), (46,000 per channel). 
All are connected to an isolated defect (or to dead columns).
\item[$\bullet$] There are around 1000 defects per CCD channel, that is, about 10$^5$ for the whole camera.
\item[$\bullet$] Many types of defects are found. Some with a few tagged pixels and some with hundreds of tagged pixels.
For instance, the defects a, c and d, circled in Fig. \ref{fig:fig2}, are examined in Fig. \ref{fig:fig8} : $\underline{a}$ is circular 
with no absorption, $\underline{c}$ is strongly absorbing with a complex shape, $\underline{d}$ is slightly absorbing with no interference 
rings. In addition, $\underline{b}$ is a single absorbing pixel surrounded by 8 pixels at half level, that is, a dead pixel.
\item[$\bullet$] The defect distribution is sufficiently sparse to separate individual defects and to build a comprehensive catalog.
\item[$\bullet$] For each tagged pixel, we measure a significant $\overrightarrow{\nabla}$ vector. Therefore a given defect is characterized by a vector field. 
\end{itemize}

\begin{figure}
\centering
\includegraphics[width=1.0\linewidth]{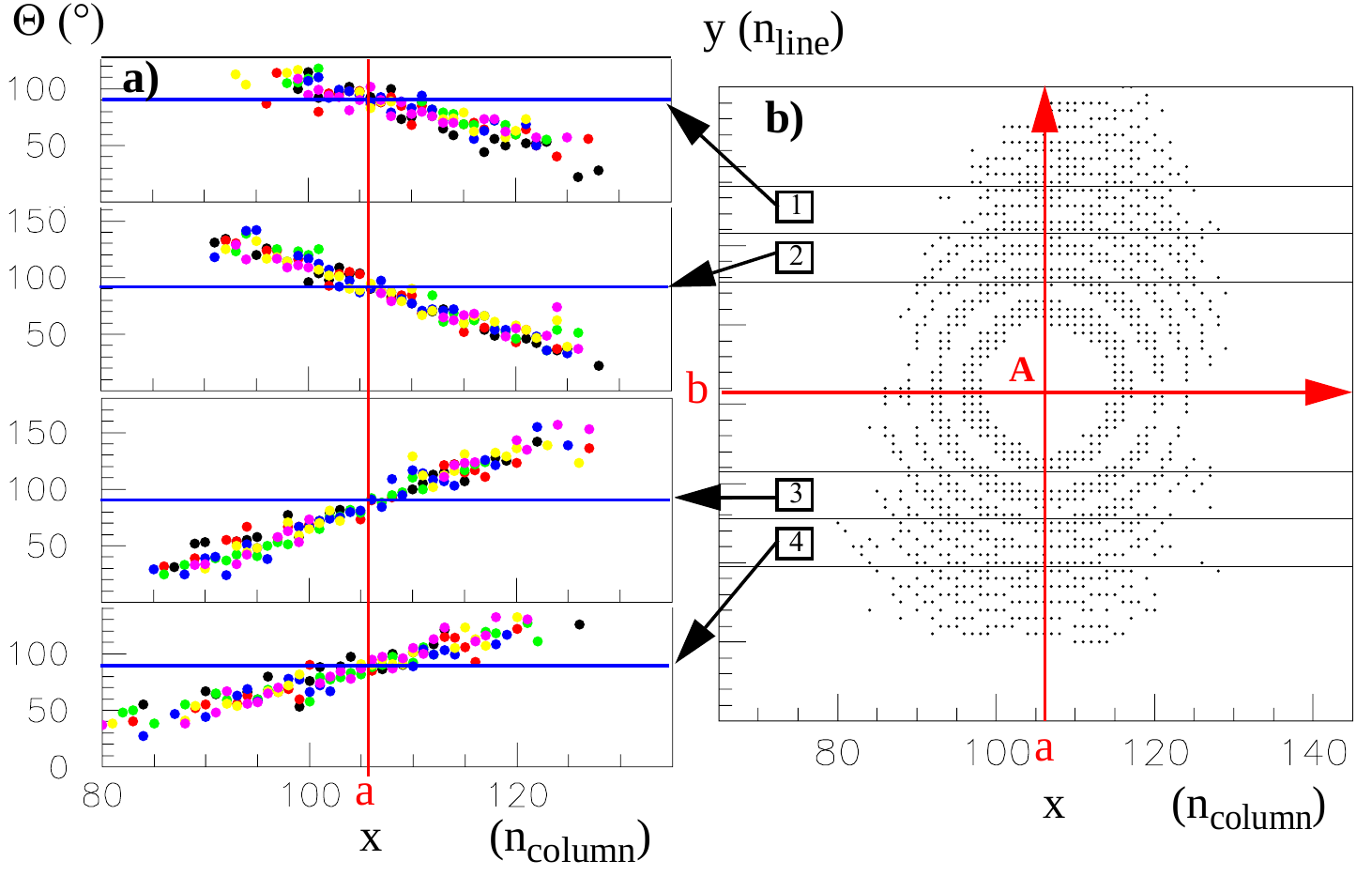}
\caption{$\bf a)$ Gradient field pattern around a defect A is characterized by a regression analysis of angle $\Theta$ 
(cotg$\Theta$=$\mathrm {\nabla}$.Ox) versus pixel x in each CCD line y. The regression equation is x=a+(b-y)cotg$\Theta$. $\bf b)$ 
Four horizontal slices of six lines each are drawn corresponding to groups of lines in a). The field lines converge on the point A 
{x=a, y=b}.
\label{fig:fig9}}
\end{figure}

Figure \ref{fig:fig9} shows how the analysis of the vector field transforms the cloud of pixels produced by tagging into field lines 
and a center of curvature $\underline{A}$. It defines piecewise the phase contours by joining some concentric arcs of circle. This method was 
adapted for contours that are more circular at the periphery of a cluster than in the central region where the center of gravity 
$\underline{A}$ of the defect is. A regression analysis fits a common center of curvature A(a,b) for parallel phase contours defined by 
the property of field lines: (x-a)sin$\Theta$ + (y-b)cos$\Theta$=0.

\section{High-precision CCD electronics}
\label{sec:4}

The aim of this paper is to track the precision limit of photometry as it is applied in astronomy. The basic concept is to count photons 
along the three quantum steps: LED emission, telescope transmission and CCD absorption. The second step is the only one that uses optics. It 
has been treated above in Sect.\ref{sec:2}. The remaining analysis in Sect.\ref{sec:5} is pure statistics. However, the precision of the 
third step, photon counting by Megacam, is currently limited by the instability of the electronics. We observe hourly fluctuations of the 
gain in a 0.8\% interval and of the pedestal around 0.1 mV (15 adu). The resulting problems are mitigated for astronomers by the 
empirical subtraction of local sky background and by the comparison to local reference stars, but maybe not as well as needed. In 
Sect.\ref{sec:41} we expose the mitigating techniques developed especially for Sndice images that reach a 1-30 ppm fluctuation 
range. Then in Sect.\ref{sec:42} we describe why we are confident that better CCD readout electronics might yield all the precision 
needed without resorting to mitigation.

\subsection{Mitigation of Megacam electronics problems}
\label{sec:41}

Megacam electronics problems (cf. \citealt{ref8}) are expressed by a variation of the pedestal and gain constants in Eq. (\ref{eq:2}) 
depending on the image number and by fluctuations during image readout. The pedestal fluctuations, which yield horizontal and vertical 
lines in Fig.\ref{fig:fig3}, are controlled using the $\nabla_x$ and $\nabla_y$ filters, which suppress these lines selectively. The gain 
fluctuations are controlled using the stability of the Sndice light source. This could yield a vicious circle because we need gain 
corrections to yield CCD data and vice versa. To break the loop, we introduce the concept of a flux$\times$efficiency$\times$gain (FEG) 
scale. It is based on the fact that the fraction of the total number of photons impinging a half CCD, noted b$_k$ in Eq. (\ref{eq:2}), is 
constant within its $\approx$3$\times$10$^{-6}$ statistical fluctuation. The observed fluctuation of the CCD count is due to a common 
multiplicative factor $\psi_{k,I}$ of the flux, the gain, or the efficiency in the Eq. (\ref{eq:2}). We slightly transform this equation 
by replacing the global flux $\Phi_I$ by the FEG average $\psi_{I}$ :
\begin{equation}
 \psi_{I}=<\psi_{k,I}>_{k=1,72}=\Phi_I \times <\epsilon_k \times g_{k,I}>_{k=1,72} \quad\quad  .
\label{eq:7}
\end{equation}

The average of the gains of the 72 channels is an order of magnitude more stable than the gain of one channel ($\approx$0.02\% rms versus 
$\approx$0.2\% rms). This fluctuation is on the same order of magnitude as the effect of the LED thermal fluctuations (0.1$\degree$C) on 
the flux $\Phi_I$ . Both contribute equally to the FEG fluctuations. Using only ADC counts, we cannot distinguish between them. We 
determine $\psi_{I}$ at $\approx$3$\times$10$^{-5}$ precision, an order of magnitude better than each of its components $\psi_{k,I}$. The 
relative gain parameter of each channel is the real one divided by the 72-channel gain average. In practice, we fix a reference image and 
fix the gains for the other images relative to this reference. With this FEG scale, we mimic what would be done with an ideal 
electronics: we would check that the gains in each image are compatible with those of the reference image at a 3$\times$10$^{-5}$ level. 
Then their average could be tested at the next order of precision, that is, 3$\times$10$^{-6}$, which is the limiting precision of the photon 
statistics in one channel (i.e., one half CCD). This precision is reached after mitigation in the remaining study, except in 
Sect.\ref{sec:56}, where the FEG mitigating method is replaced by the use of the real gain of each channel determined by the photon 
noise in the actual frame.

\subsection{Making high-precision CCD electronics}
\label{sec:42}

We claim that making an ideal CCD readout electronics is feasible rather easily with modern technology. We base this claim on our 
experience with a large electronic system in the H1 experiment (\citet{ref7}) calibrated at a 30 ppm level for 15 years and on a R\&D on 
Megacam electronics (\citet{ref5} and \citet{ref6}) reaching a 0.2 e equivalent noise charge. High-precision electronics would open a 
wide range of applications to the methods developed in this paper. One example is the preventive maintenance of a telescope using a 
measurement much more sensitive than the usual scientific requirements (i.e., detecting problems before they hurt). Another example is the 
creation of photometric standards and the photometric calibration of any instrument (not only telescope) at the ppm level. This paper also 
prooves that for the low light fluxes of astronomy, cooled CCDs are the best photometric calibrators\footnote{better than cooled 
large area photodiodes used by Sndice, which are in turn better than the NIST warm photodiodes (calibrated in the ill-defined 
photovoltaic mode)}.

\section{Coherent illumination calibration at the quantum precision limit}
\label{sec:5}

We have regrouped in Sect.\ref{sec:51} the mathematical methods used when comparing CCD images. Readers interested in bare results could 
skip it. However, this paragraph is needed to understand why the precision of our regression analyses is so good. These methods are used 
in Sect.\ref{sec:52} and Sect.\ref{sec:55} to measure the single parameter that defines a particular image inside a sequence: the total 
number of photons emitted by the LED during the exposure.  In Sect.\ref{sec:54} we show how the stability of the led position during a 
sequence has to be controled. Finally, in Sect.\ref{sec:56} we check for two image sequences (8.5 and 24 billions of pixels, respectively) 
that the content of each pixel in each image is entirely defined by quantum mechanics and photon statistics.

\subsection{Mathematical properties of the statistical distributions of the four-vector $\{{\mathrm \alpha, \beta, \gamma, \delta}\}$} 
\label{sec:51}

Our measurement model is Eq. (\ref{eq:2}). Groups of four pixels are replaced by the four filters\footnote{introduced as 
differential operators in Sect.\ref{sec:24}, they act as filters on a Fresnel spectrum}, according to 
Eq. (\ref{eq:5}). The uncorrelated noise, which is the quadratic average of the noises coming from the four individual pixels, is the 
same for each filter. In contrast the correlated electronic noise and the led-jitter noise are different. As shown in Sect.\ref{sec:2}, 
the $\delta$ filter suppresses the pedestal, the correlated gain fluctuations, and the led jitter at a 10$^{-6}$ level. 
The $\beta$ and $\gamma$ filters suppress correlated noises down to a few 10$^{-4}$ level. The $\alpha$ filter keep most correlated 
noises that are at a few 10$^{-3}$ level and the led jitter noise. (We recall that according to the Parseval theorem, signal and noise power 
are globally conserved in the Fourier transform and in the orthogonal change from the pixel basis to the filter basis in Eq. 
(\ref{eq:5})).

The next step is to consider the 1.18 million four-vectors inside a given half-CCD k of a given image $I$ as the successive occurrence of 
four random variables $\alpha_{k,I},~ \beta_{k,I},~ \gamma_{k,I}$, and~ $\delta_{k,I}$, themselves the components of a random four-vector 
$\Xi_{k,I}$. Equation (\ref{eq:2}) yields that the expected value of $\Xi_{k,I}$ is the result of applying the filters to 
the WP signal seen by one given half-CCD. The sequence of 1.18 million four-vectors almost perfectly simulates those taken by a 
multivariate Gaussian variable whose distribution is represented by Eq. (\ref{eq:9}):
\begin{equation}
\begin{split} 
&\langle\Xi_{k,I}\rangle = [\Psi_{k,I}\quad 0\quad 0\quad 0]\quad\quad\;
 \Xi_{k,I}' = \Xi_{k,I} -  \langle\Xi_{k,I}\rangle \quad  \\
\label{eq:9}
&\langle\Xi_{k,I}' \times \Tilde{\Xi}_{k,I}' \rangle =\Psi^2_{k, I}  
\begin{bmatrix}
 \sigma^2_{\alpha_{k}}  &  0  & 0  & 0  \\
 0  &   \sigma^2_{\beta_{k}}  & 0  & 0  \\
 0  &  0 & \sigma^2_{\gamma_{k}}   & 0  \\
 0  &  0 & 0  & \sigma^2_{\delta_{k}}   \\
\end{bmatrix} \quad 
 \sigma^2_{\beta_{k}} =  \sigma^2_{\gamma_{k}} \quad\quad .   
\end{split} 
\end{equation}
The relations in Eq. (\ref{eq:9}) have all been verified. First, the mean of $\alpha$ has been defined in Eq. (\ref{eq:7}).  The three 
other means are null within a fraction of an adu. This property is explained theoretically using the Fourier analysis of the WP signal as reported
in Sect.\ref{sec:2}. Second, the covariance matrices are diagonal due to the algebraic properties of the WP phase contours and 
because of the rotation invariance. The values of $\sigma_{\alpha}$, $\sigma_{\beta}$, $\sigma_{\gamma}$, and $\sigma_{\delta}$ are directly related 
to the Fresnel spectra seen in Fig.\ref{fig:fig4_b} ($\sigma_{\alpha}\approx$3.5\%, $\sigma_{\beta}=\sigma_{\gamma}\approx$0.9\%, and 
$\sigma_{\delta}\approx$0.4\%). They do not depend on the flux of image $I$ and not much on the channel number $k$. Therefore we sometimes dropped the 
indices $k$ and $I$ in their expression. Moreover, <$\beta$|$\gamma$>=0 and $\sigma_{\beta}=\sigma_{\gamma}$ because of rotation invariance. After 
associating a 4$\times$4 diagonal Gaussian with each CCD channel, we added to it the diagonal Gaussian noise in Eq. (\ref{eq:10}).
\begin{equation} 
\begin{split} \label{eq:10} &\langle \delta \Xi_{k,I}' \times \delta \Tilde{\Xi}_{k,I}' \rangle = 
\begin{bmatrix}
 \varsigma^2_{\alpha_{k,I}}  &  0  & 0  & 0  \\
 0  &   \varsigma^2_{\beta_{k,I}}  & 0  & 0  \\
 0  &  0 & \varsigma^2_{\gamma_{k,I}}   & 0  \\
 0  &  0 & 0  & \varsigma^2_{\delta_{k,I}}   \\
\end{bmatrix} \quad\quad .
\end{split}
\end{equation}
The 72 four-vector variables $\Xi_{k,I}'+\delta \Xi_{k,I}'$ are also centered Gaussians. Their means are null and their variances, measured 
independently for each image, are the raw $\Psi\sigma$ flux estimators. They are plotted in Fig.\ref{fig:fig17}.b) for the level ramp and in 
Fig.\ref{fig:fig19_a} for the flux ramp (after dividing the expression by $\Psi_{k,I}^2$ and averaging all channels). 
To extract the pure WP signal from the noise, we compared different images two by two.  Extending Eq. (\ref{eq:9}) to all pairs of images 
$I_1$ and $I_2$ leads to Eq. (\ref{eq:11}).

\begin{align}\label{eq:11}
&\langle\Xi_{k,I_1}' \times \Tilde{\Xi}_{k,I_2}' \rangle = 
 \Psi_{k,I_1} \Psi_{k,I_2}
\begin{bmatrix}
\sigma^2_{\alpha_{k}}  &   \eta_{x_k} \Delta_{x_{I_1 \rightarrow I_2}} &   \eta_{y_k} \Delta_{y_{I_1 \rightarrow I_2}}  & 0  \\
 \eta_{x_k} \Delta_{x_{I_2 \rightarrow I_1}}  &   \sigma^2_{\beta_{k}}  & 0 & 0 \\
 \eta_{y_k} \Delta_{y_{I_2 \rightarrow I_1}}   &  0 & \sigma^2_{\gamma_{k}}   & 0 \\
0  &  0 & 0  & \sigma^2_{\delta_{k}}   \\
\end{bmatrix} 
\\ \notag
 &\Delta_{x_{I_1 \rightarrow I_2}} = -\Delta_{x_{I_2 \rightarrow I_1}} \quad\quad\quad 
\Delta_{y_{I_1 \rightarrow I_2}} = -\Delta_{y_{I_2 \rightarrow I_1}} \quad\quad .
\end{align}
The compact matrix form of Eqs. (\ref{eq:9}), (\ref{eq:10}) and (\ref{eq:11}) hides a great complexity. For example, the total number of 
variables N$_t$ = 4$\times$72$\times$N$_I$ is 20160 for the sequence of N$_I$=70 images in the flux ramp. This yields 20,160 diagonal terms and 
2,812,320 pairs of non-diagonal terms of interest. This is the number of terms that we analyse in Sect.\ref{sec:52}. When we restrict the 
distribution of these enormous Gaussian variables to some components $x$ and $y$, it yields a bivariate Gaussian law with a two$\times$two 
covariance matrix C$_{xy}$. The C$_{xy}$ matrix is written conventionally with the two marginal variances $\sigma_x^2$ and $\sigma_y^2$ on the 
diagonal and a non-diagonal term $\rho \sigma_x \sigma_y$ ($\rho$ being the correlation coefficient). The additivity of covariance matrices 
allows us to add the noise in Eq. (\ref{eq:10}) to the WP signal of Eq. (\ref{eq:11}). This yields the following equation:
\begin{equation}
\begin{split}
C_{\alpha_{k,I_1} \alpha_{k,I_2}} = &
\underbrace{\sigma^2_{\alpha_{k}}
\begin{bmatrix}
\Psi^2_{k, I_1}  &  \Psi_{k, I_1} \Psi_{k, I_2}   \\
\Psi_{k, I_1} \Psi_{k, I_2}  &   \Psi^2_{k, I_2}   \\ 
\end{bmatrix}
}_{WP}
+
\underbrace{\begin{bmatrix}
\varsigma^2_{\alpha_{k, I_1}}  &   0 \\
0 & \varsigma^2_{\alpha_{k, I_2}}   \\ 
\end{bmatrix}}_{Noise}
\\
= &
\begin{bmatrix}
\sigma^2_{x} &  \rho\sigma_{x}\sigma_{y} \\
\rho\sigma_{x}\sigma_{y}  &    \sigma^2_{y} \\ 
\end{bmatrix} \quad\quad .
\end{split}
 \label{eq:12}
\end{equation}

In Sect.\ref{sec:52} we assume for the bivariate Gaussian distribution of two $\alpha_{k,I}$ variables a common representation of the regression 
analysis: $x$ is the marginal variable, $y$ the conditional variable, and the three parameters are $\sigma_{x}$, $\sigma_{y/x}$, and the slope 
$a_{y/x}$. Classical formulas\footnote{Formulas and their application to our problem are found in \citet{ref8}, Appendix C.} that relate the two 
parametrizations of the regression analysis are used in Sect.\ref{sec:52} to evaluate the difference between the slope $a_{y/x}$ of the 
regression line and the gain ratio of WP signal $\Psi_y/\Psi_x$ ($x=\alpha_{k, I_1}$ ; $y=\alpha_{k,I_2}$). This difference 
$D=2(\sigma_{y/x}/\sigma_{\alpha})^2$ $\approx$1\% is small for the $\alpha$ variables at the reference flux, supporting the choice of a 
regression estimator for the gain ratio in this case. But D is large for the other three variables $\beta$, $\gamma$, and $\delta$, imposing 
another type of noise estimator, the variance of $\Delta \delta$ (in which $\delta$ by may be replaced by $\beta$ or $\gamma$):
\begin{equation}
\Delta \delta_{k,cur} = \delta_{k,cur} - \xi_{k,cur}\delta_{k,ref}
\qquad
\xi_{k,cur} = \Psi_{k,cur}/\Psi_{k,ref} \quad\quad .
\label{eq:17}
\end{equation}
This linear combination of the current and the reference images eliminates the WP signal on a pixel-by-pixel basis and yields a pure noise variable.
Its mean is null and its variance, using Eq. (\ref{eq:12}) is: 
\begin{equation}
\langle \Delta \delta^2_{k,I} \rangle / \Psi^2_{k,I} =
\varsigma^2_{k,I} / \Psi^2_{k,I}  + \varsigma^2_{k,ref} / \Psi^2_{k,ref} =
S_k(\Phi_I) + S_k(\Phi_{ref})  \quad\quad .
\label{eq:18}
\end{equation}
The identification of the square of $\varsigma_{\delta}/\alpha$ with the so-called statistical factor S($\Phi$) is a key of the analysis of 
uncorrelated noise in Sect.\ref{sec:56}.

In summary, the four-vector Gaussian model yields four mean estimators and a variance matrix (four variances and six covariances) for one image. 
Three means and four covariances are null. We are left with the mean of $\alpha$ and four variances used in the following as five redundant flux 
estimators, plus two covariances used as led motion estimators. For a sequence of images we extract four sequences of noise estimators based on 
the variance of the flux-weighted difference of two images (one for each filter).

\subsection{High-precision flux ratio estimates}
\label{sec:52}

The algorithm estimating the flux ratio of the two images $I_1$ (reference) and $I_2$ (current) was introduced by \cite{ref12}. Its 
principle, which is illustrated in Fig.\ref{fig:fig10}, considers that the photon distribution is multinomial (not Gaussian). The ratio of 
the FEG variables $\alpha_{k,I}$ in a given four-pixel matches the ratio of exposure duration, which is about 5/6 in this example. We reconstructed 
the joint probability distribution as a product of the marginal distribution of the reference variable $\alpha_{k,ref}$ (left) and the 
conditional probability of the current variable $\alpha_{k,cur}$(right). The joint distribution has three properties:

\begin{figure}
\centering
\includegraphics[width=1.0\linewidth]{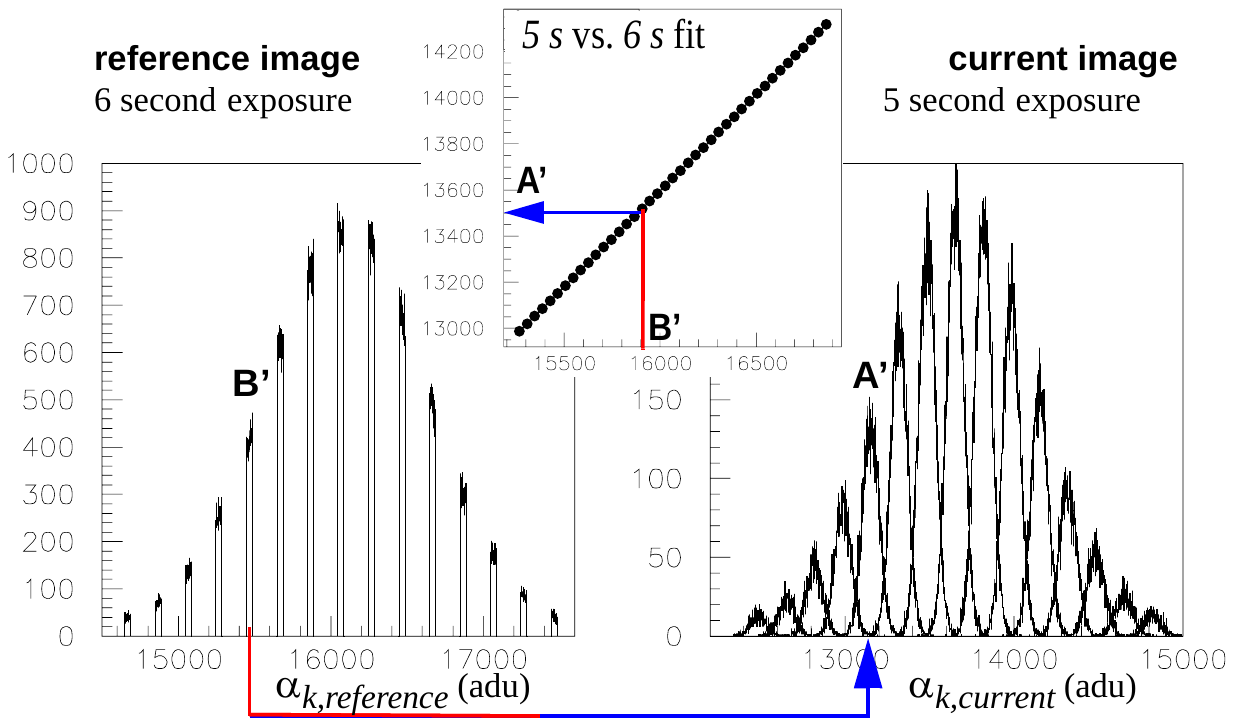}
\caption{ $\bf Left$ the histogram of $\alpha$ in the reference image subdivided in slices 40-adu wide. $\bf Right$ the histogram of the 
projection of each reference slice in any current image is a Gaussian. Only one slice for every five is represented. $\bf Top$(inset) the 
means of the current slices are fit as a linear function of the means of the reference slices. The distribution of the residuals is shown in 
Fig.\ref{fig:fig11}.}
 \label{fig:fig10}
\end{figure}

	a) The marginal distribution is gaussian:

A Gaussian fit\footnote{The truncated Gaussian fit is good within $\pm 2 \sigma_{\alpha}$ ; 4 out of 72 channels have non-Gaussian tails due to stains 
on the mirror (see \citealt{ref8}, fig.12).} yields  <$\alpha_{k,ref}$>= $\Psi_{k,ref}$ and $\sigma(\alpha_{k,ref}) = \Psi_{k,ref} \sigma_{\alpha}$. 

	b) The regression curve is a straight line (see inset of Fig.\ref{fig:fig10}): \begin{equation}
<\alpha_{k,cur} / \alpha_{k,ref}>= \Psi_{k,cur} +   a_{y/x}(\alpha_{k,ref}-\Psi_{k,ref}) \quad\quad .
 \label{eq:13}
\end{equation}
In each bin of the $\alpha_{k,ref}$ histogram, we fit a Gaussian on the $\alpha_{k,cur}$ distribution and hence give a value of the 
conditional mean $<\alpha_{k,cur}/\alpha_{k,ref}>$ and of the conditional standard deviation $\sigma(\alpha_{k,cur}/\alpha_{k,ref})$. The quality 
of the fit of Eq. (\ref{eq:13}) is excellent, as shown in Fig.\ref{fig:fig11}-a, where individual errors bars are the Gaussian width divided 
by the root of event number, or more conservatively, in Fig.\ref{fig:fig11}-b by the width (0.4 adu rms) of the distribution of residuals. It determines 
$\Psi_{k,cur}$ with a 0.06 adu (rms) point precision (4$\times$10$^{-6}$). Particular care is taken for such high-precision point measurements 
involving a small fraction of an adu. They are valid only as representing an average of the digital sampling of a continuous analog variable over a 
wide ADC range. In this example, where $\alpha_{k,cur}$ is sampled within a 14000$\pm$750 adu interval, the precision of 
the fit of the slope $a_{y/x}$ is -1.7$\times$10$^{-4}$.

\begin{figure}
\centering
\includegraphics[width=1.0\linewidth]{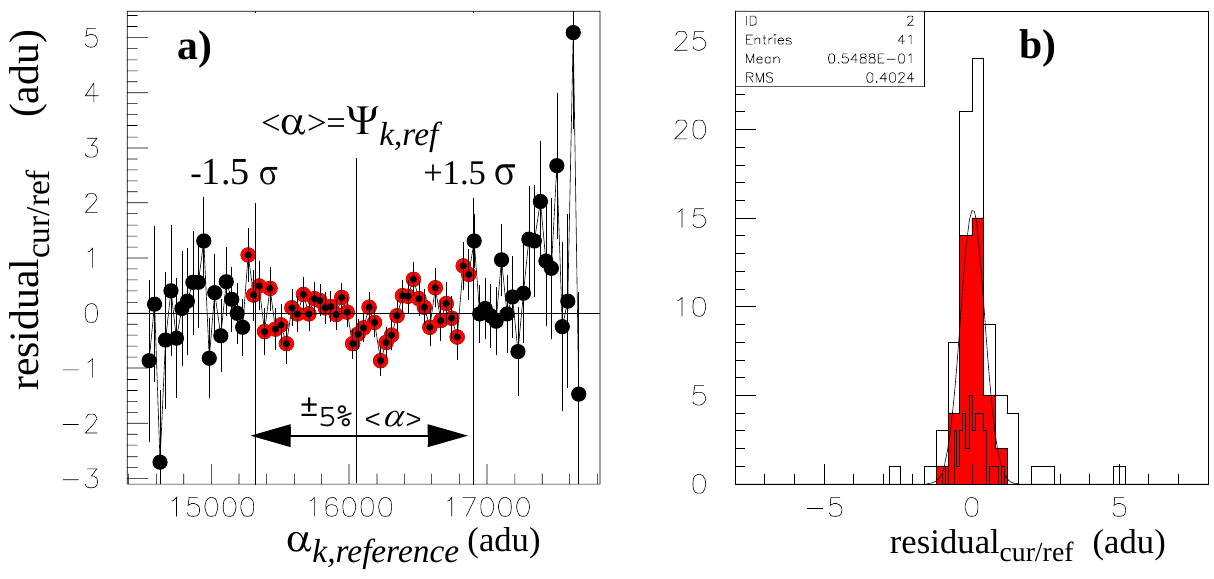}
\caption{Residuals of the linear fit of Fig.\ref{fig:fig10}:  $\bf a)$ as a function of $\alpha_{k,reference}$; $\bf b)$ as a histogram
(0.4 adu rms). In red: points included in the fit ($\alpha \in [\langle \alpha \rangle \pm $1.5$\sigma ]$)}.
 \label{fig:fig11}
\end{figure}
\begin{figure}
\centering
\includegraphics[width=1.0\linewidth]{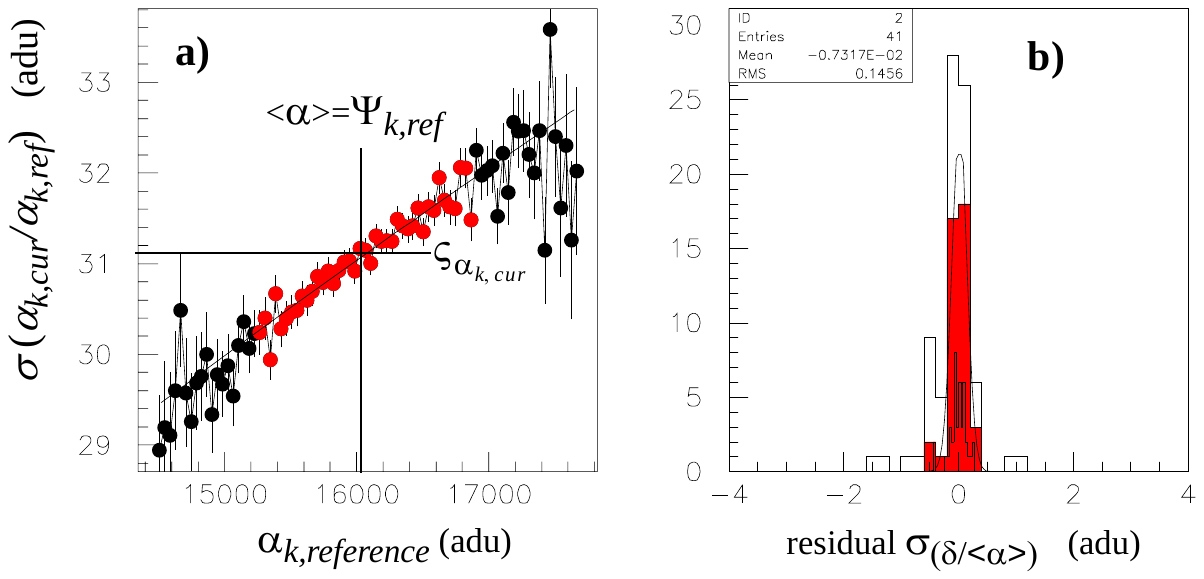}
\caption{$\bf a)$ Fit of the width of the joint distribution as a function of $\alpha_{k,reference}$ in the $\{\langle \alpha \rangle \pm $1.5
$\sigma \}$ interval. $\bf b)$ Histogram of the residuals of the previous fit, fitted by a Gaussian (0.14 adu rms).
 \label{fig:fig12}}
\end{figure}


	c) The conditional standard deviation $\sigma(\alpha_{k,cur}/\alpha_{k,ref})$ varies as a square root of the flux (because of the 
multinomial law of photon counts). 
We fit a polynomial on the data, as shown in Fig.\ref{fig:fig12}-a. Its first-order linear approximation is:\begin{equation}
\sigma(\alpha_{k,cur}/\alpha_{k,ref}) =
\sigma_{y/x} + b_{y/x} \times (\alpha_{k,ref} - \Psi_{k,ref}) \quad\quad .
 \label{eq:14}
\end{equation}
The central value $\varsigma_{k,cur}=\sigma_{y/x}$ will take the place of the constant value defined for a Gaussian. The point 
precision in this example is excellent (0.02 adu $\approx\Psi_{k,ref}\times$10$^{-6}$). The precision on $b_{y/x}$ is 0.023 adu/adu. 
This process of extrapolation at the central reference flux $\Psi=\Psi_{k,ref}$ of the flux-dependent quantities f($\Psi$), used in Eq. (\ref{eq:13}) 
and (\ref{eq:14}), is applied systematically to all other variables.

\subsection{Determining the LED jitter using the $\alpha'/\beta$ and $\alpha'/\gamma$ correlation}
\label{sec:54}

\begin{figure}
\centering
\includegraphics[width=1.0\linewidth]{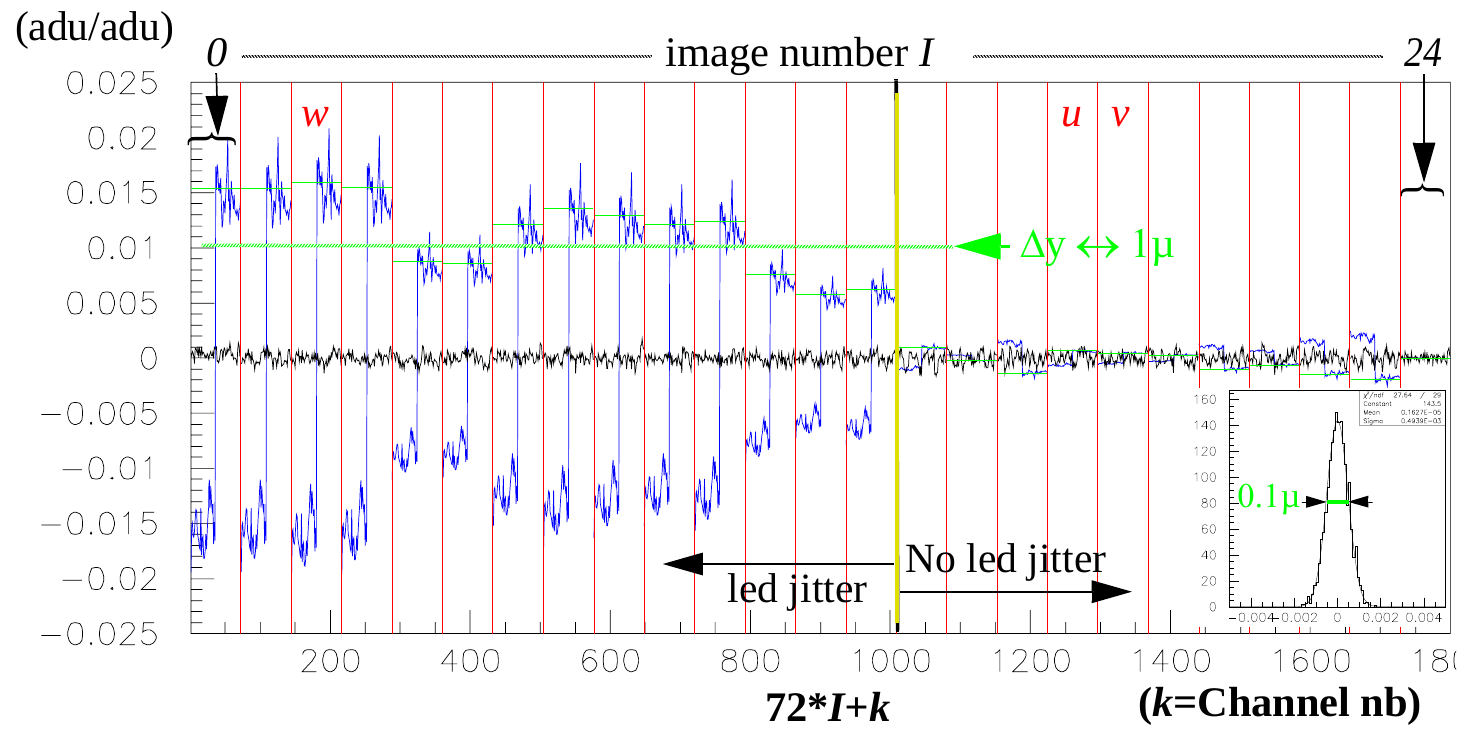}
\caption{In blue we show the signal of $\gamma$ vs. $\alpha$ correlation as a function of channel $k$ and image $I$ (noted $\eta_k \Delta y_I$ , with 
$\Delta y_I$ fixed to 0 for $I$=24). In black we plot the residual of the fit of this signal with one $\eta_k$ parameter per channel and one 
$\Delta y_I$ per image. Images $I$=2,17,18 were called $w$, $u$, $v$ in the spectral studies of Sect.\ref{sec:2}. $\Delta y$ is calibrated by 
comparison with $\Delta y_{w\rightarrow u}$. The inset shows a Gaussian fit of the residuals (1 point/channel/image) with a 5.10$^{-4}$ rms, 
yielding $\Delta y_{k}$ = 0.05 $\mu$ or <$\Delta y_{k}$> = 0.006 $\mu$. The green line indicate the effect of a 1$\mu$ LED drift.
 \label{fig:fig14}}
\end{figure}

Two correlation terms, <$\beta$|$\alpha$> or <$\gamma$|$\alpha$>, appear in the covariance matrix of Eq. (\ref{eq:11}), while they are null in 
the autocovariance matrix of Eq. (\ref{eq:9}). An intuitive explanation of this puzzle is found in \cite{ref8}. The point of interest here is 
that the non-diagonal matrix elements noted $\Psi_{k,I_1}\times\Psi_{k,I_2}\times\eta_{y_k}\times\Delta_{y_{1 \rightarrow 2}}$ are a very 
sensitive probe of the LED jitter projected on y axis (idem for x). LED jitter is the only source of noise found in the optical signal in addition
to the photon noise. The <$\gamma$|$\alpha$> terms for each CCD channel k (0 $\le$ k $\le$ 71) in the sequence of $N_I$=25 images at constant flux 
level yields a $N_I\times N_I$ matrix. In Fig.\ref{fig:fig14}, we only keep the last row ($I_1$=24, $I_2$=0 to 24) of the matrix, but we repeat 
the operation for the 72 channels. The raw data (in blue) are the slopes $a_{\gamma/\alpha}$ of the regression fit in Eq. (\ref{eq:13}). They are 
ordered by time (that is, by image number $I$) and by electronic channel number $k$. Here $\alpha_{24}$ is the marginal variable and 
$\gamma_0,...,\gamma_{24}$ are the conditional variables. The reason for taking the reference image from among the last eleven images in 
Fig.\ref{fig:fig14} is obvious: it belongs to a group of images ($I$=14,...,24) in which led jitter is minimal.

The result of a complementary method is shown in Fig.\ref{fig:fig14}. It is a principal component analysis that fits the 1800 raw data using 72 
$\epsilon_k \eta_k$ and 25 $\Delta y_I$ parameters ($a_{\gamma / \alpha}= \epsilon_k \eta_k \times \Delta y_I$ ; <$\epsilon_k \eta_k$>=1). The 
index $\epsilon_k$=$\pm$1 is introduced to take the up/down orientation of the CCD readout within the focal plane into account. It explains the 
characteristic data pattern: negative for the first 36 and positive for the last 36 channels, or vice versa. The fit values of the vertical 
displacement $\Delta y_I$ are drawn in green and the residuals of the fit in black. The calibration of the $\Delta y$ scale was made using 
Fig.\ref{fig:fig6}, where the displacement of the LED between image $w$ ($I$=2 and $\Delta y$=0.016) and image $u$ ($I$=17 and $\Delta y$=0.0005) 
is estimated at 1.4 $\mu$. This yields a 1\% per micron calibration ratio of the a$_{\gamma/\alpha}$ slope change per LED displacement. The 
distribution of residuals in the inset of Fig.\ref{fig:fig14} displays a 0.05\% Gaussian width, that is, a sensitivity for the LED position given by 
one channel equal to 0.05 $\mu$ rms. The average sensitivity for all 72 channels is 0.006 $\mu$, that is, a mean angular position of the LED defined 
at 0.4 nrad.

We conclude this study by observing that we are fortunate to have a rather good mechanical stability of the telescope illumination system, because 
there was no provision for this effect during the construction of Sndice. We did not yet perform a study of the mechanical stability, but we note 
that the flux ramp run during two hours was affected by no $\delta y$ displacement and only one significant $\delta x$ displacement. This is used 
in the next paragraph to obtain a full flux ramp unaffected by led jitter.

\subsection{Determining the fluxes for a sequence of images}
\label{sec:55}

\begin{figure}
\subfigure{\includegraphics[width=0.8\linewidth]{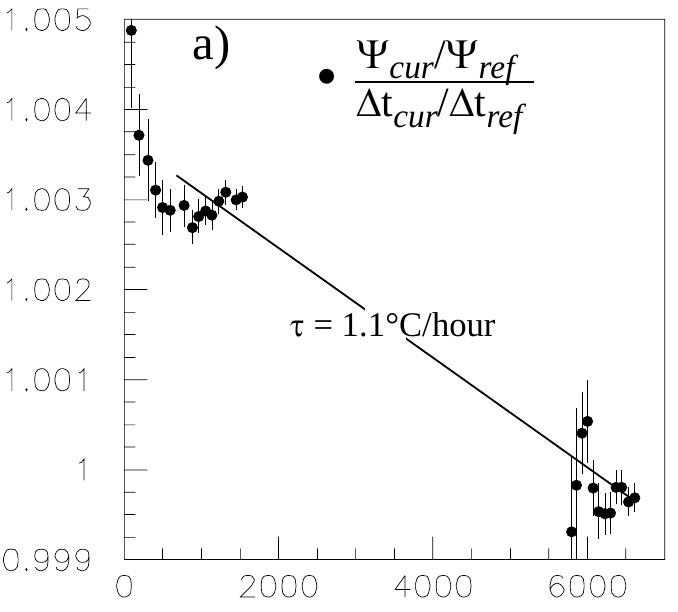}}
\subfigure{\includegraphics[width=0.8\linewidth]{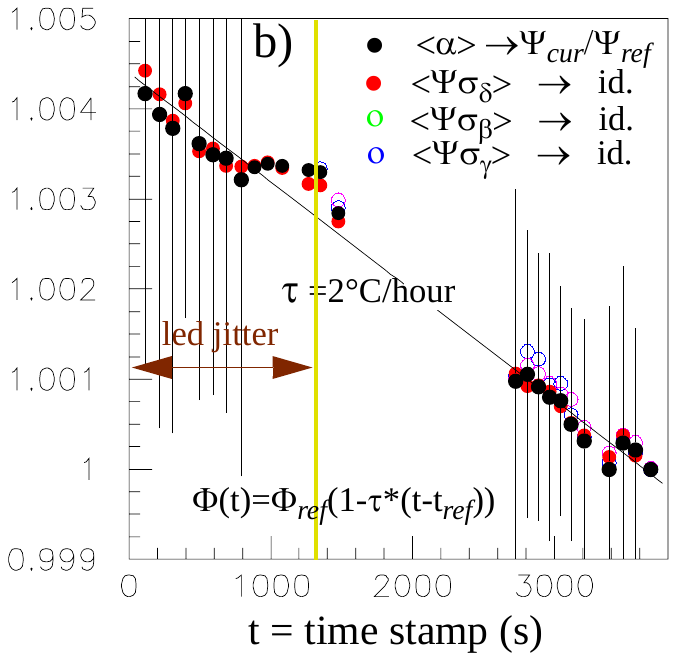}}
\caption{Effect of LED temperature on light flux (warming of CFHT dome at dawn): {\bf a)} Variable exposure (1s<$\Delta t$<8s) {\bf b)} Constant 
exposure ($\Delta t$=8s): two independent estimators <$\alpha_k$>$\approx$16000 adu (black points) and <$\Psi\sigma_\delta$>$\approx$100 adu (red 
points) agree within 0.8$\times$10$^{-4}$ rms. Deviation from linearity is 1.8$\times$10$^{-4}$ rms. The precision on <$\Psi\sigma_\delta$> is 
$\approx$0.008 adu, i.e., 0.6$\times$10$^{-6}$. Error bars cover gain spread before averaging. The two other estimators <$\Psi\sigma_\beta$> and 
<$\Psi\sigma_\gamma$> (green and blue) are more sensitive to LED jitter.
 \label{fig:fig17}}
\end{figure}

The integrated flux $\Phi_I$ emitted by an LED is a product of LED current, exposure time, and temperature terms (cf. Eq. (\ref{eq:2})). It is 
measured by the LED electronics. In Fig.\ref{fig:fig17} we represent the trend due to the linear temperature variation of  
1$\degree$C and 2$\degree$C per hour (due to the warming of the CFHT dome after dawn). Under these conditions, the test-bench calibration of 
Sndice tells us that the precision is limited to a few 10$^{-4}$ and could be improved to a few 10$^{-5}$ by monitoring the LED current and the 
temperature (\citealt{ref11}, §4)\footnote{There was no monitoring of the LED current and temperature during Megacam data taking.}. In the constant 
level run the exposure time was kept constant by means of an electronic LED shutter defined at a 0.3 $\mu$s time resolution. In the flux ramp the 
exposure time was varied using the megacam shutter (1ms resolution).

Alternatively, we measured the mean flux absorbed in the CCDs. The linear fit of Eq. (\ref{eq:13}) yields for each channel $k$ a constant term 
$\Psi_{k,cur}/\Psi_{k,ref}$ and a slope term $a_{y/x}$, with a statistical precision of 4$\times$10$^{-6}$ and 
1.7$\times$10$^{-5}$/(0.1$\times\Psi_{k,ref}$), respectively. LED jitter has no effect on the constant term of the fit. The large error bars seen in 
Fig.\ref{fig:fig17}.b) show the spread of the gain fluctuations in the 72 channel data. The mitigation method described in Sect.\ref{sec:41} 
reduces the gain fluctuations ($\delta g_{k,cur/ref}$ $\approx$1.5$\times$10$^{-3}$ rms) and yields an average FEG flux ratio (black points),
\begin{equation}
\Psi_{cur}/\Psi_{ref} =\langle \Psi_{k,cur}/\Psi_{k,ref}\rangle_k
= (1+ \langle \delta g_{k,cur/ref}\rangle_k) \times
\Phi_{cur}/\Phi_{ref} \quad\quad .
\label{eq:16}
\end{equation}
The deviation from the linear trend is 1.8$\times$10$^{-4}$ rms. It is compatible with the averaging of 72 channels ($\delta g$/$\sqrt{72}$ 
=1.5$\times$10$^{-3}$/$\sqrt{72}$). Thermal fluctuations of LED, around 0.1$\degree$C per minute, have comparable effects.

\begin{figure}
\centering
\includegraphics[width=1.0\linewidth]{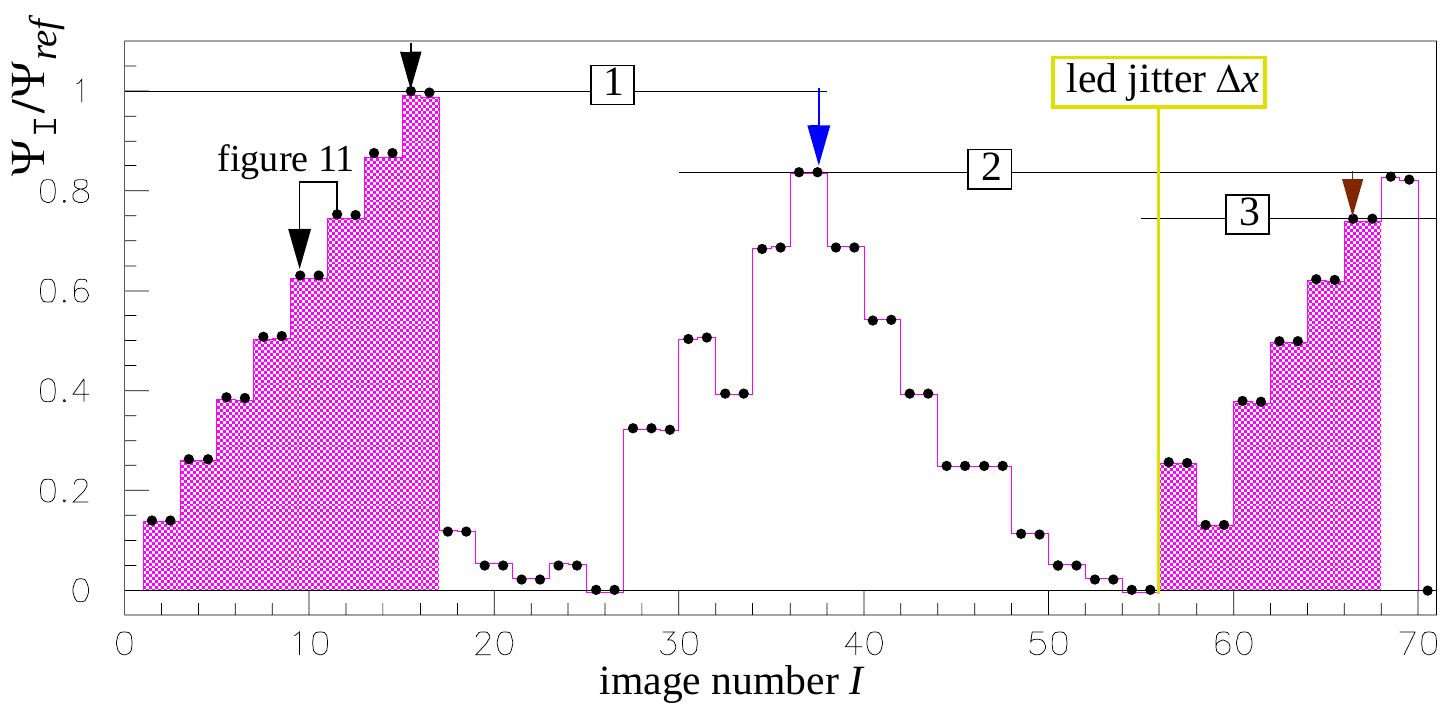}
\caption{Setting up the relative fluxes within a sequence of 70 images: $\bf 1)$ Three sequences, covered by horizontal lines, are built around three 
reference images (black, blue, and brown arrows). Gain$\times$fluxes ratios are determined as in Fig. \ref{fig:fig10} ($I_{cur}$=9, $I_{ref}$=11). 
$\bf 2)$ The reference image of sequence 2 is measured relative to reference 1 and reference 3 relative to reference 2. By transitivity all 
fluxes are related. $\bf 3)$ The two overlap regions 1 over 2 and 2 over 3 yield a set of double determinations. The relative fluxes of all 70 
images agree within 3.10$^{-5}$ rms. They give the relations flux vs. exposure time (shaded area: 28 images at a common LED current) and flux 
vs. LED current (constant exposure time). The only significant LED jitter occurs between images 55 and 56.
 \label{fig:fig15}}
\end{figure}

For the variable exposure run, the 70 images in Fig. \ref{fig:fig15} yield a point representing $\Psi_{cur}/\Psi_{ref}$ the ratio of its averaged 
FEG flux over that of a reference image. Integrated fluxes are varied by two different means : \uline{exposure time} using shutter speed (magenta 
shade) or \uline{LED current} (plain). As a precaution, because the long periods at low flux destroy the continuity of high-precision data, we 
took three reference images marked by vertical arrows (one for each peak of flux). To reconnect the results based on different 
references, we measured the relation between each pair of reference images and checked the transitivity of the flux ratio measurements. The 
relative flux precision that we obtained at highest flux is $\approx$3$\times$10$^{-5}$ rms. The mechanical shutter yields the error bars 
($\delta$($\Delta$t)= 1ms) seen in Fig. \ref{fig:fig17}-a). Clearly, the electronic shutter is prefered. The <$\alpha$> flux ratio 
estimator $\Psi_{cur}/\Psi_{ref}$ measured so far reaches a precision of around 10$^{-4}$ after mitigation of the electronics errors. It is 
essentially a measurement of the flux of specular light.

Four other measurements, $\Psi_{k,I}(\sigma_{\alpha}, \sigma_{\beta}, \sigma_{\gamma}, \sigma_{\delta})$, the square root of the covariances in 
Eq. (\ref{eq:9}), yields four completely independent estimates of the flux based on the diffused light ($\approx$10\% of specular light). The 
application of these covariance estimators provides a positive test of the WP model with the spectacular precision shown in Fig. 
\ref{fig:fig17}-b. The $\Psi_{k,I}\sigma_{\delta}$ estimate yields the red points superimposed on the black ones. There is a 0.8$\times$10$^{-4}$ 
rms agreement between the two types of estimators. The agreement is better than the 1.8$\times$10$^{-4}$ precision resulting from averaging the 
gains in Eq. (\ref{eq:16}). This is explained by considering that both types of estimators are based on the same FEG scale and not on the real 
flux scale. The 0.8$\times$10$^{-4}$ precision on $\sigma_{\delta}$ corresponds to a 0.008 adu precision on the pixel counts. This result, 
relative to the average pixel content of 16000 adu, entails a remarkable precision on the WP hessian signal width $\Psi\sigma_{\delta}$ of 
0.5$\times$10$^{-6}$ rms, almost at the statistical precision limit of the 10$^{13}$ photons. The two other quantities shown in the figure -$\Psi 
\sigma_{\beta}, \Psi\sigma_{\gamma}$- yield similar results, but the analysis is complicated by the introduction of the LED jitter noise, which 
adds up quadratically to the WP signal dispersion. The limiting precision for the WP estimators is set by the photon noise. The filtering of the 
low spatial frequencies suppresses the effect of electronic bugs.

\subsection{ $\alpha$, $\beta$, $\gamma$, and $\delta$ noise estimators}
\label{sec:56}

\begin{figure}
\centering
\includegraphics[width=0.7\linewidth]{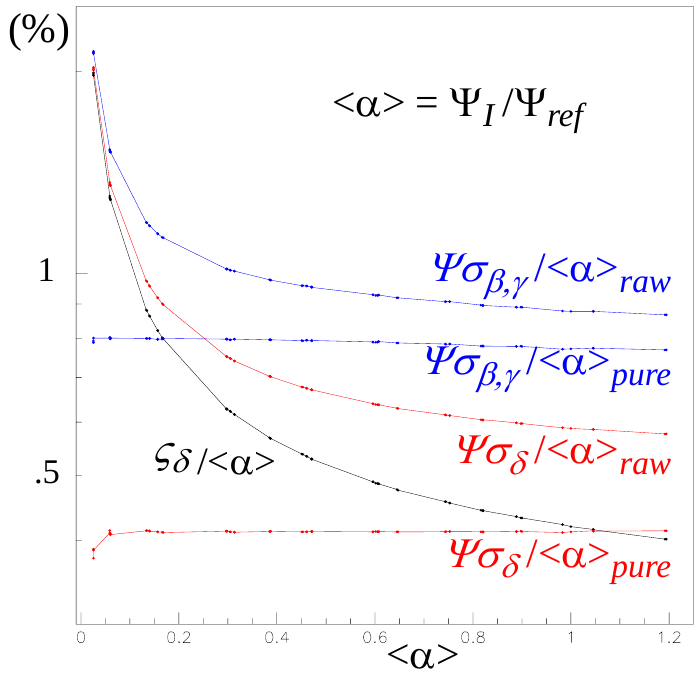}
\caption{ Pure noise $\varsigma_{\delta}$ is extracted by subtracting the raw reference image (with its relative weight) from all other raw 
images. Then pure WP signals $\Psi \sigma_\beta$, $\Psi \sigma_\gamma$, and $\Psi \sigma_\delta$ are extracted from raw images by subtracting noise 
$\varsigma_{\delta}$. Check: $\sigma_\beta, \sigma_\gamma$, and $\sigma_\delta$ are constant and $\sigma_\beta =\sigma_\gamma$ (superimposed).
\label{fig:fig19_a}}
\end{figure}

The raw variance of a filter content in Eq. (\ref{eq:12}) is the sum of the WP variance $\Psi_{k,I}\sigma_{\delta}$ and the noise variance 
$\varsigma_{\delta}$.  Figure \ref{fig:fig19_a} reproduces the variances of three filters in a relative form (divided by the FEG flux <$\alpha$>).  
This figure is one half of the consistency check of our model. It shows on a very broad dynamical range that the interference pattern is proportional 
to the flux (because it is defined by the probability density of the wave packet). The second half of the demonstration is contained in the 
analysis of the noise variance as a function of flux (Fig.\ref{fig:fig21}), because it demonstrates that in most of the range the noise is 
dominated by the photon statistics.

Without too many technical details, we report that we applied the slicing method and the fits shown in Fig.\ref{fig:fig11} and 
Fig.\ref{fig:fig12} to the distributions ($\delta_{k,cur}$, $\alpha_{ref}'$) and ($\Delta\delta_{k,cur}$, $\alpha_{ref}'$). The variances of 
$\delta_{k,cur}$ and $\Delta\delta_{k,cur}$ (Eq. \ref{eq:17}) yield the estimates for the raw WP signal and for the pure noise, respectively. 
Both are shown in Fig.\ref{fig:fig19_a} for three filters.

\begin{figure}
\subfigure{\includegraphics[width=0.9\linewidth]{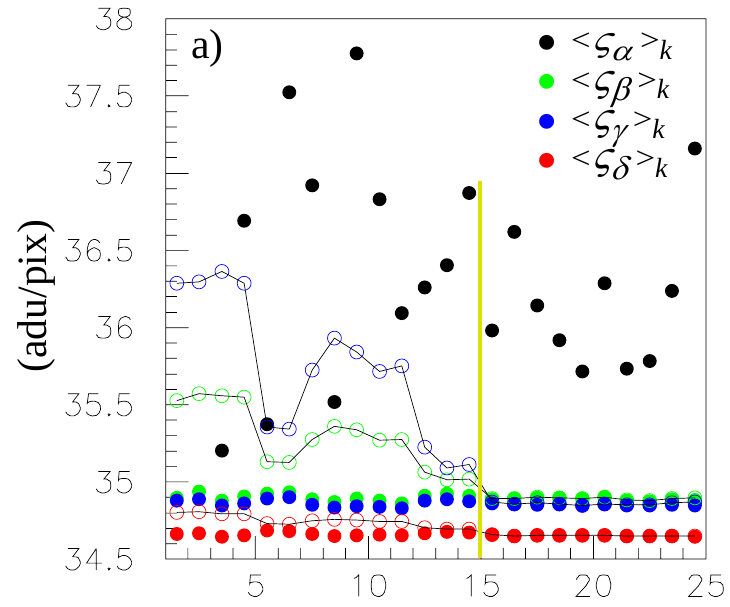}}
\subfigure{\includegraphics[width=0.9\linewidth]{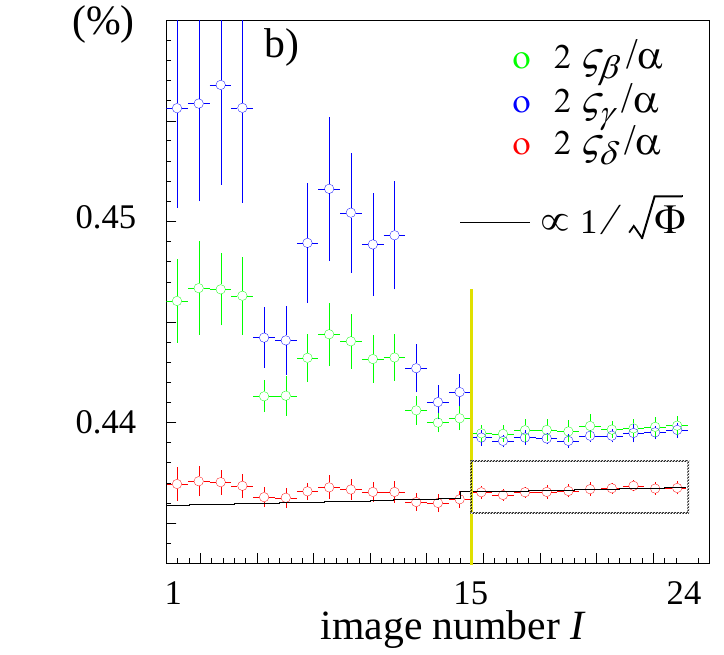}}
\caption{{\bf a)} Fluctuations $\varsigma_{\alpha,...,\delta}$ (rms) of $\alpha,...,\delta$ between the last image $I$=25 and any other one $I$=1,24 
in temporal order (72 channels average). Open circles represent raw data and full circles data corrected for LED jitter. Variable $\alpha$ senses all 
noise sources; $\beta$ ($\gamma$) suppresses the line (column) correlated electronic noise and the LED jitter along x (y) axes; $\delta$ 
suppresses all correlated noises and LED jitter. {\bf b)} Relative fluctuations 2$\times$<$\varsigma_{\beta(,\gamma,\delta)} / \alpha$> are 
compared to the prediction. The continuous line representing the prediction $\propto$1/$\sqrt{\Phi(t)}$ uses the flux $\Phi$(t) drawn in Fig. 
\ref{fig:fig17}-b. Error bars are given by the $\Psi_k$ dispersion (k=1,72). The individual channel precision is 5$\times$10$^{-6}$, the 72 channel 
average precision is 0.7$\times$10$^{-6}$.
\label{fig:fig18}}
\end{figure}

We used the constant level run as a benchmark for the high-precision noise estimators. A good summary of the study is seen in Fig. 
\ref{fig:fig18}. It represents two versions of the same four ($\alpha$, $\beta$, $\gamma$, and $\delta$) noise variance estimators. For the upper one 
(Fig.\ref{fig:fig18}.a), mitigation yields only one average adu count per image per filter proportional to the average adu count of 
the reference image. A global LED jitter noise correction was applied using the parameter $\Delta y$ from Fig. \ref{fig:fig14} (open circles 
before and full circles after correction). In the lower one (Fig.\ref{fig:fig18}.b), there are 72 data per image (one relative noise per 
channel). Relative noise is not affected by gain fluctuations (cancelled between the numerator and the denominator). Data are represented in the 
plot by their mean and rms. The precision is sufficient to fit the flux dependence of the noise (proportional to 1/$\sqrt{\Phi_I}$). The 
comparison of the four filters after LED jitter correction gives the size of the correlated fluctuations among four neighboring pixels. Fully 
correlated fluctuations such as pedestal or gain fluctuations are seen by the $\varsigma_{\alpha}$ variable. Their effect is in a 0.5-3 adu 
range. The line-to-line (or column-to-column) fluctuations sensed by $\varsigma_{\beta}$ (or $\varsigma_{\gamma}$) yield a 0.25 adu effect. The 
fourth variable $\varsigma_{\delta}$ serves as a pure sample of uncorrelated noise to be used for a fine study of the photon noise on the 
whole flux range covered by the 70 images flux ramp.

The three uncorrelated random processes affecting the WP signal have a different flux dependence: the pedestal noise is constant, the gain noise is 
proportional to the flux, and the photon noise to the square root of the flux. The variances were added to constitute what is classically called 
the statistical factor (Eq. (\ref{eq:19})). We could take into account the LED jitter variance in the statistical factor, but we do not need it
because the flux ramp is divided into two sequences with no internal LED jitter, 
\begin{equation}
\begin{split}
S_k(\Phi) = &\; (\varsigma_k(\Phi)/\Phi)^2
= (A_k + B_k\Phi + C_k \Phi^2)/ \Phi^2\\ 
S_k(\Xi) = &\; C_k + B_k \Xi + A_{k}\Xi^2 \qquad
\Xi = \Phi_{ref} / \Phi \quad\quad .
\end{split}
\label{eq:19}
\end{equation}
The link between the variance of the noise variable $\Delta \delta_{k,I}$ in Eq. (\ref{eq:17}) and the statistical factor has been given in Eq. 
(\ref{eq:18}), which sums the statistical factors of the current image and the reference image. This eliminates not only the WP signal, but also 
the fluctuation of gains of both current and reference images, which are the root of our electronic problems. The change of variable from the 
flux $\Phi$ to its inverse $\Xi=\Phi_{ref}/\Phi$ in Eq. (\ref{eq:19}) transforms S($\Phi$) into a second-degree polynomial in $\Xi$. The 
photo-electron noise is in the $B_k\Xi$ term and the pedestal noise in $A_{k}\Xi^2$.

\begin{figure}
\centering
\includegraphics[width=1.0\linewidth]{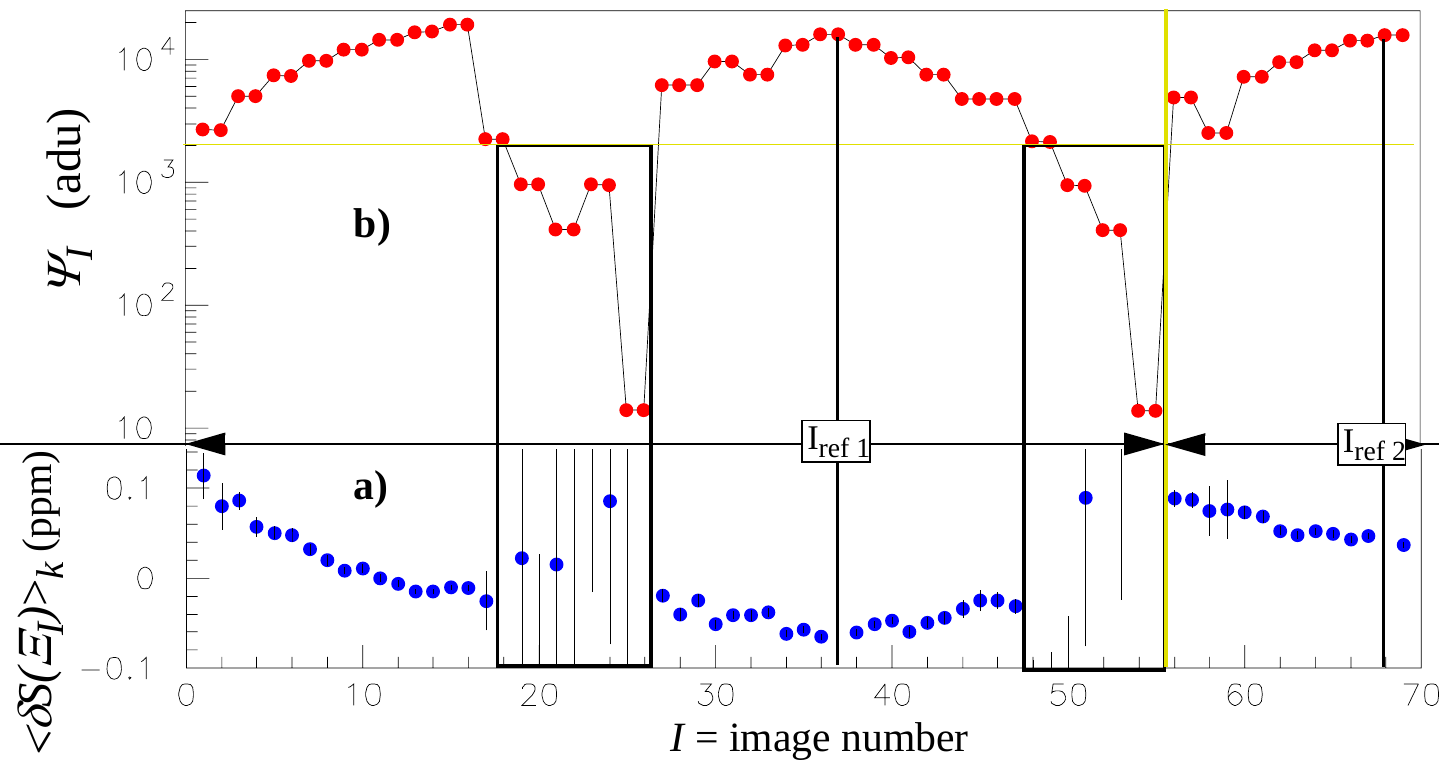}
\caption{{\bf a)} Bias <$\delta S_k(\Xi_I)$>$_k$ (in blue) drift continuously with time. It equally affects all 72 residuals of the fit 
S($\Psi$) vs. $\Psi$. When corrected, the width of the residual distribution is reduced to its photon statistics value (Fig. \ref{fig:fig21}-a). This 
correction is equivalent to a modification of the flux scale $\Psi_I$ $\rightarrow$ $\Phi_I$. {\bf b)} The flux ramp sequence $\Psi_I$ (red), taken 
from Fig. \ref{fig:fig15}, is correlated to the dispersion of biases (error bars in Fig. \ref{fig:fig20}-a). For $\Psi_I$ <2000 adu (black boxes) 
the dispersion of the pedestals dominates the dispersion of the photon number.
 \label{fig:fig20}}
\end{figure}

\begin{figure}
\centering
\includegraphics[width=1.0\linewidth]{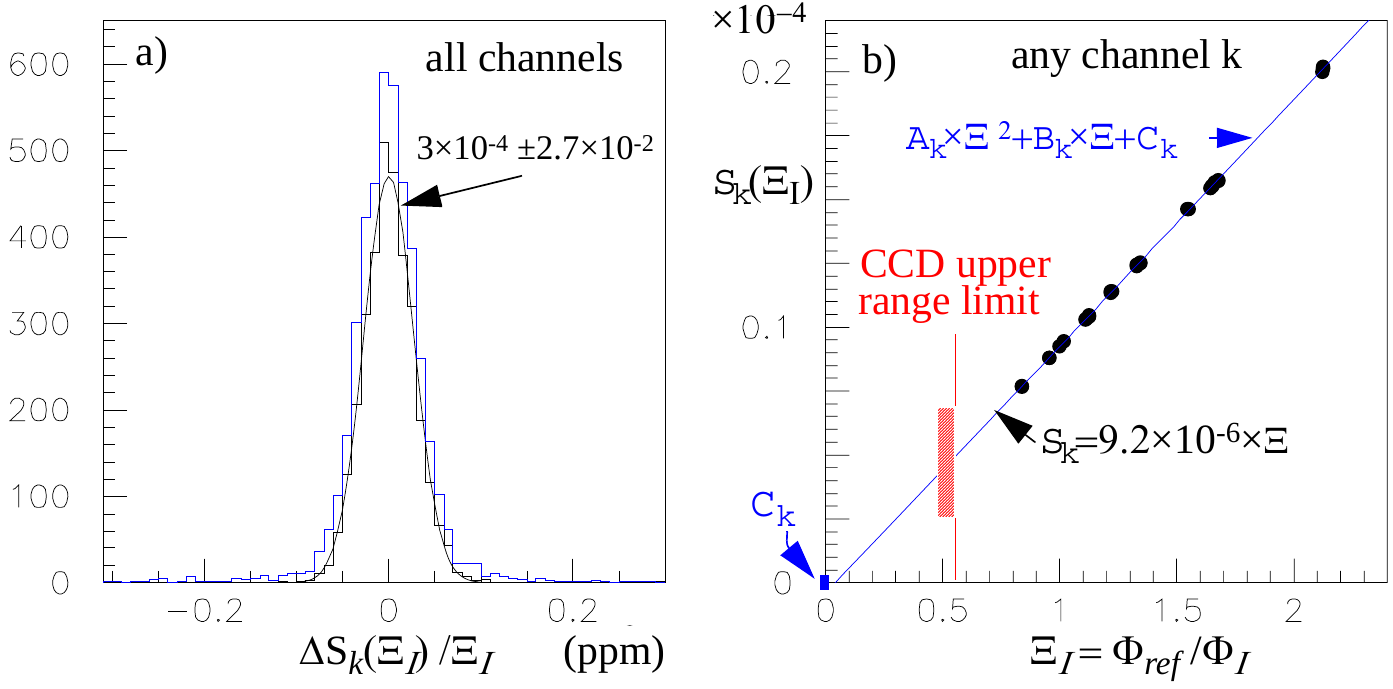}
\caption{{\bf a)} Residuals of $S_k(\Xi_I)$ vs. $\Xi_I$ fits (k=1, 72, I=1;70). In black we plot data and Gaussian fit for $\Psi_I$>2000 adu 
with a width =2.7$\times$10$^{-8}$ rms; in blue we show the complete data. {\bf b)} Extrapolation at infinite flux for any channel k yields a 
negligible value of C$_k$=S$_k$(0).
 \label{fig:fig21}}
\end{figure}

Figure \ref{fig:fig21}.b shows one of the 72 curves representing the S($\Xi_{k,I}$) vs. $\Xi_{k,I}$ data and their fit by a second-degree 
polynomial on a flux range of two orders of magnitude. The second-degree term is visible only when extending the flux range down, from hundreds to 
tens of photo-electrons per pixel. For each point of a S$_k(\Xi)$ curve a gain fluctuation does not alter the ordinate S$_k(\Xi_I)$ but shift the 
abscissa $\Xi_I$. The shift of $\Xi_I$ from $\Psi_{ref}$/$\Psi_I$ to $\Phi_{ref}$/$\Phi_I$ is common to all the 72 channels of an image. Figure 
\ref{fig:fig20}.a displays the residuals of the 72 linear fits of S$_k(\Psi_I)$ vs. $\Psi_I$, which contain the common mode effect of the $\Phi_I 
- \Psi_I$ shift in addition to the random noise. This effect is statistically significant, therefore we corrected for it. The correction reduces the 
dispersion of residuals seen in Fig. \ref{fig:fig21}.a by a factor two for all channels. This amounts to replacing the FEG flux $\Psi_I$ by an FE 
flux $\Phi_I$ (or to correct the fluctuation of the average gain assuming that the average efficiency is constant). Figure \ref{fig:fig20}.a 
represents a continuous drift of the average gain with time independently of the flux. The overall distribution of final residuals, shown in Fig. 
\ref{fig:fig21}-a, is an unbiased Gaussian with a 2.7$\times$10$^{-8}$ rms. For channel 72, whose S($\Xi$) vs $\Xi$ fit is given in Fig. 
\ref{fig:fig21}-b, this entails a 0.30\% Gaussian width of $\Delta S_k/S_k$. Using the S=($\varsigma_{\delta}/\alpha$)$^2$ relation, 
$\Delta\varsigma_{\delta}/\alpha$ =0.5$\times$($\Delta$S/S)$\times \varsigma_{\delta}/\alpha$ = 4.5$\times$10$^{-6}$ rms (at reference flux 
$\Xi$=1).\footnote{Another way to quote the precision is $\Delta(\varsigma_{\delta}/\varsigma_{\delta})$= 0.5$\times$($\Delta$S/S)= 0.15\%} This 
is the number expected from a pure photo-electron statistical noise, which proves that there is no other unknown or uncorrected systematic 
fluctuation in the CCD measurement.

In addition to photon noise, our random noise model in Eq. (\ref{eq:19}) contains the two auxiliary terms $C_k$ and $A_k$. In Eq. (\ref{eq:18}), 
our noise estimator, $S_k(\Phi_{ref})$ is a constant added to $C_k$. In practice, we fit a second-degree polynomial $P_2(\Xi)$ to the data and 
evaluate it at $\Xi$=1. This yielded $P_2$(1)=2$\times S_k(\Phi_{ref})$. The constant $S_k(\Phi_{ref})$ =$P_2$(1)/2 was subtracted from the data. 
Then we repeated the fit on these reduced data. The new constant term is the real $C_k$ seen in Fig. \ref{fig:fig21}-b. For all channels $C_k$ is 
null within a good approximation: there is no need to envisage a noise component other than photon statistics in the wide range above 2000 adu.
The coefficient $A_k$ includes the Johnson noise of the amplifier and fluctuations of the pedestal, reaching a few adus. We see in Fig. 
\ref{fig:fig20}-b that, first in the 2000 -20000 adu range (above the yellow line) the $A_k$ term is too small compared to the signal to be 
sensed and a linear fit is perfect, then in the 200-2000 adu range (inside rectangles) $A_k$ is needed and the dispersion of the residuals (error 
bars) increases as a result of the pedestal fluctuations, and finally in the range below 200 adu (not sampled in the ramp), pedestals should be processed 
differently.

In the present section we emphasized the importance of an accurate $\Phi$ scale for the photon noise ramp. Previously in Sect.\ref{sec:55}, we 
developped an accurate $\Psi$ scale needed for the WP signal ramp. This does not set the two scales on the same footing. The $\Phi$ scale is 
already at its theoretical precision limit of 1.8$\times$10$^{-4}$ because of the photon statistics, while the $\Psi$ scale precision is limited 
by the poor stability of electronics, also at 1.8$\times$10$^{-4}$. $\Psi$ scale could be improved by two orders of magnitude by using high 
precision electronics, to reach its photon statistical limit, and both scale could be reconciled at a common value better than 10$^{-5}$.

\section{Conclusions and perspectives}
\label{sec:conclusion}

We have shown in this paper that there is no other limit for a photometry based on the Megacam camera than the statistical fluctuations of the 
photon count in any CCD area, set in our case to below 1 ppm per exposure for the whole image. The proof, using the properties of some difference 
operators applied to the photon field, is indirect because of the defects of the electronics. The single photo-electron response is calibrated by 
statistics and the integrated flux is measured for each exposure using the total response of the whole detector or a part of it. We might call 
this type of photometry self-consistent or self-calibrated.

If the LED flux or the CCD flux are deduced from electronic readings alone, the precision is limited by the stability and the calibration of 
electronics. After optimizing the electronics this precision limit should be around 20-30 ppm. In practice, with either Sndice or Megacam 
electronics, the current precision is degraded to 100-200 ppm. We showed how electronics might be optimized to suppress these 
practical limitations. We call this type of photometry electrically calibrated. The precision of an optimized electric calibration could be 
maintained for years. It surpasses the best photometric results obtained using stable stellar sources. Electric calibration allows comparing 
different exposures of a varying light source or monitoring the evolution of a detector with a constant LED source, while self-calibrated fluxes 
are used to compare even more precisely two sources with a common detector or two detectors with a common source (e.g., for calibration transfer). 
Based on these examples, we can build a large set of direct illumination calibration applications. A third type of calibration, the absolute 
calibration, is done in SNDICE by a common practice method using a NIST calibrated photodiode in a test bench. In this case, we would speak of 
accuracy instead of precision. We did not discuss accuracy because the absolute calibration procedure mentioned above is not sufficiently 
reliable. First, because it does not consider either the angular dependence and the map of the detector quantum efficiencies or the emission 
pattern of the light sources. Second, because in the light of our study of cooled large area photodiodes for other Sndice publications, we cannot 
take the accuracy of the photovoltaic quantum efficiency used by NIST  for granted as the reference photodiode yield.

At this point, we could have concluded the review of the instrumental results obtained from our Sndice-on-Megacam data by observing that we had 
reached a level of precision better by two order of magnitude than best astronomical photometry and that we measure the effect of diffuse reflection 
on the mirror that is not seen by other means with the same precision.

However, it was more fruitful to take a new point of view: to consider the interference patterns that we called the WP signal like a signal to be 
studied instead of a noise to hide (as calibration systems using incoherent extended illumination do). The WP signal represents 10\% of the flux 
seen in Megacam. It is stable and we measure it at an overall 10$^{-5}$ precision level. Classical signal processing methods have been applied to 
the WP signal. They produced simple and useful results. In frequency space, the WP spectrum separates the specular and diffuse components of 
light. It measures the effect of the photon propagator in free space and yields optical surface quality estimators. It also maps the defects of 
optical surfaces individually. As compared with holographic or phase-contrast systems with similar abilities, we have had a large-scale 
high-performance system for free already built in the camera.

In direct space, we have used and developed pixel difference operators (PDE) with extremely useful results. In particular, the extraction of 
photon statistical noise is performed by four independent operators that first define the angular position of an LED in the telescope frame at a 0.4 
nrad rms precision and then yield four noise estimators at better than 1 ppm.

The perspective of entering a new territory of high-precision photometry should be seriously considered. The first steps might  easily be to 
build dedicated photometric systems and improve commercial components such as LEDs for our purposes.

\begin{acknowledgements}

This work incorporates a number of important contributions to the subject that must be quoted and for which I express all my gratitude. First of 
all is the Megacam camera including its E2V CCDs, which represents an epochal realization. I am grateful to Pierre Borgeaud, who introduced us to 
the Megacam electronics and helped starting the electronic developments described in Claire Juramy's thesis. In this thesis we also found the first 
proposition of a direct illumination calibration of a telescope with LEDs that we developed together in a later article. Reynald Pain together 
with the CFHT scientific and technical team transformed this concept into the SNDICE project. Kyan Schahmaneche has led the realization of Sndice 
with its associated calibration bench in record time and its installation in Hawaii. Kyan and Augustin Guyonnet did the calibration of Sndice 
on the bench. Augustin's thesis started the study of the high-precision methods presented here. Sndice calibration tools were developed and 
advanced in an article by Nicolas Regnault. He applied these tools with Marc Betoule to the calibration of the SNLS experiment. Last but not 
least, we benefited from illuminating discussions with Pierre Astier.

\end{acknowledgements}

\def\aap{{\em A\&A}}
\def\apj{ApJ}
\def\apjl{ApJ Lett.}
\def\apjs{ApJ Supp.}
\def\aj{AJ}
\def\prd{Phys. Rev. D}

\bibliographystyle{aa}
\bibliography{biblio}

\end{document}